# The Same Analysis Approach: Practical protection against the pitfalls of novel neuroimaging analysis methods

Kai Görgen[a], Martin N. Hebart[bc], Carsten Allefeld[a]*, John-Dylan Haynes[ade]*

[a] Charité – Universitätsmedizin Berlin, corporate member of Freie Universität Berlin, Humboldt-Universität zu Berlin, and Berlin Institute of Health (BIH); Bernstein Center for Computational Neuroscience, Berlin Center for Advanced Neuroimaging, Department of Neurology, and Excellence Cluster NeuroCure; 10117 Berlin, Germany

[b] Department of Systems Neuroscience, University Medical Center Hamburg-Eppendorf, Martinistr. 52, 20251 Hamburg, Germany

[c] Section on Learning and Plasticity, Laboratory of Brain & Cognition, National Institute of Mental Health, National Institutes of Health, Bethesda MD, USA

[d] Humboldt-Universität zu Berlin, Berlin School of Mind and Brain and Institute of Psychology; 10099 Berlin, Germany

[e] Technische Universität Dresden; SFB 940 Volition and Cognitive Control; 01069 Dresden, Germany

* These authors contributed equally to this work

**Correspondence:** Kai Görgen, BCCN Berlin, Philippstr. 13, Haus 6, 10115 Berlin, Germany. kai.goergen@bccn-berlin.de

# Highlights

- Traditional design principles can be unsuitable when combined with cross-validation
- This can explain both inflated accuracies and below-chance accuracies
- We propose the novel "same analysis approach" (SAA) for checking analysis pipelines
- The principle of SAA is to perform additional analyses using the same analysis
- SAA analysis should be performed on design variables, control data, and simulations

## Abstract

Standard neuroimaging data analysis based on traditional principles of experimental design, modelling, and statistical inference is increasingly complemented by novel analysis methods, driven e.g. by machine learning methods. While these novel approaches provide new insights into neuroimaging data, they often have unexpected properties, generating a growing literature on possible pitfalls. We propose to meet this challenge by adopting a habit of systematic testing of experimental design, analysis procedures, and statistical inference. Specifically, we suggest to apply the analysis method used for experimental data also to aspects of the experimental design, simulated confounds, simulated null data, and control data. We stress the importance of keeping the analysis method the same in main and test analyses, because only this way possible confounds and unexpected properties can be reliably detected and avoided. We describe and discuss this Same Analysis Approach in detail, and demonstrate it in two worked examples using multivariate decoding. With these examples, we reveal two sources of error: A mismatch between counterbalancing (crossover designs) and cross-validation which leads to systematic below-chance accuracies, and linear decoding of a nonlinear effect, a difference in variance.

## Introduction

Research practice in psychology and cognitive neuroscience has traditionally been guided by principles of experimental design and statistical analysis, much of which was pioneered by R. A. Fisher (1925, 1935). The purpose of these principles is to observe effects as clearly as possible under conditions of noisy and limited data, and to make reliable inferences about the relation between experimentally manipulated and measured variables in the presence of potentially confounding influences.

Methodological work has led to an established corpus of design principles, e.g. counterbalancing (also known as crossed or crossover design) and randomization, and statistical tests (such as *t*-test and ANOVA; Coolican, 2009; Cox and Reid, 2000). A researcher can normally apply these without extensive further checks, and their use in published work provides transparency for reviewers and readers. Cognitive neuroimaging has followed the lead and adapted these principles to the specific properties of its large, high-dimensional data sets, leading to mass-univariate GLM-based data analysis as its main workhorse (Friston et al., 1995; Holmes and Friston, 1998).

However, the complexity of neuroimaging data and the development of new theoretical ideas about neural processing have motivated a wealth of alternative analysis approaches, foremost among them multivariate pattern analysis (MVPA; Haxby et al., 2001). Driven not by standard statistical approaches, but by machine learning methods such as classification algorithms and cross-validation (Pereira et al., 2009), they significantly extended the data-analytic toolbox and made a larger variety of possible effects in neuroimaging data accessible (e.g. Kamitani & Tong, 2005; Haynes et al., 2005). The drawback of this methodological plurality is that the soundness of applied methods can no longer be judged based on an established corpus, and novel methods often prove to have unexpected properties. This is evidenced by a growing literature on possible pitfalls, pointing out e.g. that known ways to control confounds may no longer work with multivariate analysis (Todd et al., 2013), accuracies are not binomially distributed when estimated by cross-validation (Noirhomme et al., 2014; Jamalabadi et al., 2016), or a second-level *t*-test does not provide population inference if applied to information-like measures (Allefeld et al., 2016). It even applies to seemingly small extensions of established methodology, like extraction of correlations from a brain map leading to an inflated estimate (Vul et al., 2009; Kriegeskorte et al., 2009) or the use of cluster-level statistics with a threshold for which the underlying approximation might be invalid (Eklund et al., 2016).

We propose to meet these challenges to the validity of novel analysis methods by adopting a habit of systematic testing of experimental design, analysis procedures, and statistical inference, both concerning single parts and the whole analysis pipeline. The common practice of performing "control analyses" on additionally obtained data to rule out confounds (e.g. reaction time as a proxy for task difficulty) can already be seen as a limited form of such testing, but our recommendation goes beyond that: In particular, we suggest to perform analyses on aspects of the experimental design and simulated null data. A crucial point in these test analyses is that they should preserve properties of the actual pipeline as far as possible; in particular, they have to be performed using the same analysis method as the actual data analysis. For this reason, we call our proposal the *Same Analysis Approach* (SAA).

In the following we detail how to use the Same Analysis Approach, both in worked examples and in a general overview. Along the way, we reveal two possible confounds in MVPA that are not widely known in the neuroimaging community: the mismatch between a counterbalanced design and an analysis using cross-validation, and the unexpected ability of a linear classifier to "decode" differences in variance in the absence of differences in the mean. While we believe that the specific examples of errors presented in this paper are of general interest, our main aim is to highlight more generally that many types of unexpected errors can occur when there is a mismatch between design and analysis. Thus, we provide SAA as a general tool to find such errors that might affect any particular analysis pipeline in different ways.

## Example: Counterbalancing and cross-validation

A researcher intends to perform a simple neuroimaging experiment to test whether there is a difference between two experimental conditions **A** and **B**. The experiment is performed in four runs. In each run, both conditions are presented in two consecutive trials. A common confound in such a setup is the presentation order of the conditions: If **A** was always presented before **B**, it would be unclear if an *observed* difference was caused by a *true* difference between **A** and **B**, or whether it was caused by a difference between first and second trials (Figure 1a). To prevent this, the researcher employs counterbalancing (advocated by e.g. Fisher, 1935): Each experimental condition **A** and **B** is equally often presented in the first and in the second trial, i.e. equally often together with each level of the potentially confounding factor "trial order" (Figure 1b). The purpose of counterbalancing is to prevent a bias in the final analysis even if an effect of trial order was present, because if both conditions are equally often presented as trial 1 and trial 2, a systematic effect of the confound "trial order" will cancel out.

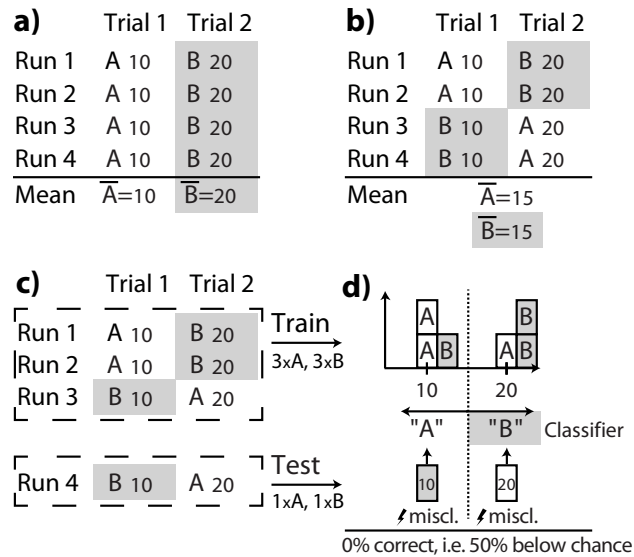

**Figure 1.** Example experiment. a,b) Experimental designs to test two experimental conditions **A** (no background) and **B** (grey background) in four runs. Small numbers (10, 20) are example data for each experimental trial. $\bar{A}, \bar{B}$ denote condition specific means. The design in panel a) cannot distinguish presentation order effects from effects of the experimental conditions, because all first/second trials are also **A**/**B** trials. The observed difference in means ($\bar{A} = 10, \bar{B} = 20$) could thus arise from a difference in either condition or trial number. In contrast, the design in panel b) controls the confound "trial order" using counterbalancing. Even if data (small numbers) would only depend on the trial number, as in this example, mean and variance of both experimental conditions are equal ($\bar{A} = \bar{B} = 15$), and thus standard statistics such as a $t$-test will not indicate a difference between the conditions: The confound control worked. c,d) The same experimental design with leave-one-run-out cross-validated classification. Panel c) shows partitioning of the data into training and test set for one cross-validation fold, and d) demonstrates that systematic misclassification of all test data arises, resulting in 0% correct predictions. Systematic misclassification will also occur in all other cross-validation folds (not shown). Thus, the intended confound control failed. Note that the reason for this systematic below-chance accuracy is neither an imbalanced number of training or test samples between conditions ($n_A = n_B = 3$ in each training set; $n_A = n_B = 1$ in each test set), nor is it specific to any particular classifier, nor to cross-validation in general. It is instead caused by a "design–analysis" mismatch between a counterbalanced design and the cross-validation scheme employed in the analysis.

## Counterbalancing works as expected for the *t*-test

To test whether counterbalancing works as expected and indeed removes the confounding effect of trial order, we can calculate what would happen if there was no difference in the experimental conditions **A** and **B**, but the data were only influenced by the confound "trial order" that is to be controlled. Figure 1a and 1b show such a situation: Neuroimaging measurements are $y = 10$ in all first trials and $y = 20$ in all second trials. Whereas the experimental design in Figure 1a does not allow to distinguish if an observed difference between **A** and **B** arises from a true difference between the condition or because **A** was always presented before **B**, the counterbalanced setup in Figure 1b does allow this distinction. Collecting the counterbalanced measurements for each condition **A** and **B** across runs yields $y_A$ = [10 10 20 20] and $y_B$ = [20 20 10 10]. Clearly, a *t*-test would not indicate a significant difference between both conditions, because the data values are identical in both. Counterbalancing therefore worked as intended: The factor "trial order" heavily confounded the data, but it had no systematic effect on the outcome of the statistical test.

**Counterbalancing does not work for leave-one-run-out cross-validation**

What happens if the counterbalanced but confounded data from Figure 1b is analysed with cross-validated classification instead of the *t*-test? Cross-validated classification is a standard MVPA method to estimate how well a classifier can learn from examples to predict ("decode") the experimental condition of independent data, and can serve to test for statistical dependency between conditions and data like the *t*-test above (Haynes and Rees, 2005; Kriegeskorte et al., 2006; Norman et al., 2006). Although cross-validated classification is typically applied to multivariate data, it can be applied to one-dimensional data equally well.[1]

In the same way as for the *t*-test above, we can check whether counterbalancing the potential confound "trial order" will also prevent unexpected effects on the outcome of the cross-validated classification analysis by assessing its performance on the confounded but counterbalanced data from Figure 1b. Selecting a specific cross-validation scheme, i.e. how to separate data into training and test sets in the different folds, is one required analysis decision. The stratified leave-one-run-out cross-validation scheme in this example is common in neuroimaging because – in contrast to data from the same run – different imaging runs can be considered approximately statistically independent. Each run contains equally many samples per class, so the cross-validation is also balanced. Because the confound "trial order" has been controlled by counterbalancing and we know that there is no effect of the experimental condition, the classifier should not be able to distinguish between the classes. In a balanced setting with two classes, "cannot distinguish" translates to a classifier that assigns conditions to data in a non-systematic fashion, leading to an expected classification performance around 50%.

---

[1] Cross-validated classification is performed by repeatedly splitting the measured data samples into independent training and test sets, inferring a relation between data and "labels" (experimental conditions) from the training set and quantifying its strength on the test set. In this example, the data of individual runs is held out in each successive split ("fold") until all data served as test data once (Figure 1c). The quality of the inferred prediction is typically measured by classification accuracy (Figure 1d). Averaging the performance measure across all folds yields an estimate of how well the classifier would perform on completely new data. This final performance estimate serves as a measure of information content and – if significantly above chance (i.e. >50% for two classes) – demonstrates a statistical relationship between experimental conditions and data. See Hebart & Baker (this issue) for a recent overview about differences and misunderstandings between MVPA and univariate analyses.

Instead, the obtained accuracy is 0% when performing the analysis, i.e. 50% below chance. This means that every single data sample was misclassified (Figure 1d). Despite the absence of a true effect, our result is worse than chance, demonstrating that in this example counterbalancing completely failed to control the confound "trial order".

**Standard control data analysis fails to detect the problem**

In an actual experiment, a researcher of course cannot know that no true effect is present in the data. Since the result *systematically* deviates from chance, they might even suspect that a true effect is present, but they might not be sure how to interpret a systematic *below*-chance accuracy (cf. Allefeld et al., 2016; Kowalczyk, 2007). Because of this, they would probably conduct further analysis to look for the source of the unexpected behaviour. For example, the researcher might perform a control analysis on reaction times that were recorded together with the neural data. The idea is that a potentially confounding variable (attention level, task difficulty, etc.) will influence both the experimental and the control data. Consequently, finding an effect on the control data would demonstrate that the main results could alternatively be explained by that confound. In this example, assume that the reaction times depend solely on trial order, just as for the neuroimaging data. For example, in first trials participants were more vigilant and responded faster, whereas in second trials their attention level was decreased and they responded more slowly (Figure 2).

Following common practice, conventional control analyses are carried out using standard univariate statistics, typically *t*-tests or *F*-tests, even if MVPA is employed for the main analysis (Todd et al., 2013; Woolgar et al., 2014). Since in this example reaction times depend solely on trial order, we already know that a *t*-test will not indicate any reaction time effect, and thus does not help explain the puzzling below-chance classification accuracy. Even though both datasets – brain responses and reaction times – are completely equivalent, the control analysis did not reveal any information about the confound that occurred in the main analysis. Thus, the control analysis failed its purpose.

|  | Run 1 | Run 2 | Run 3 | Run 4 |  |  |
|---|---|---|---|---|---|---|
|  | **A B** | **A B** | **B A** | **B A** |  |  |
| **Main data:** Neuro [a.u.] | 10 20 | 10 20 | 10 20 | 10 20 | cv-MVPA: | 0% correct ?? |
|  |  |  |  |  |  | ↑ **false security that cv-MVPA is ok** |
| **Ctrl data:** RT [ms] | 100 200 | 100 200 | 100 200 | 100 200 | t-test: | no difference |

**Figure 2.** Mismatch between main analysis and control analysis. A standard control analysis fails to detect the problem from the initial example. Upper panel: Leave-one-run-out cross-validated decoding applied to counterbalanced data without a true experimental effect but with a "trial order" confound leads to a puzzling classification accuracy of 0% (Figure 1b-d). Lower panel: A *t*-test applied to reaction times as control data does not find a difference and thereby fails to detect the problem generated by the mismatch between the counterbalanced design and the cross-validated analysis. This may lead to a false sense of certainty that results from the main analysis were not explained by a confound.

**Problem summary**

This initial example illustrates two of our main points. First, the design principle (counterbalancing) used in the experimental design comes from the established corpus but was paired with an analysis method (cross-validated classification) which does not. What the researcher overlooked here was that design principles and analysis methods do not stand on their own but work in tandem, and using another analysis method led to a "design–analysis mismatch". Second, using a standard control data analysis to diagnose the problem failed, because it did not use the same analysis method, producing a mismatch between main analysis and control analysis. Many problems that arise with the use of novel analysis methods are caused by these or similar kinds of mismatch: between design and analysis, different analysis steps, different design principles, or analysis and statistical assumptions (see below).

Note that the focus of this paper is not the specific problem outlined above, for which we could provide an explanation and solution at this point; nor is it to provide an exhaustive overview of possible errors. Instead, we introduce a general approach to diagnose problems, in form of a principled form of control analysis for any number of errors.

We now introduce this control analysis and demonstrate it on the problem of the initial example. After that, we explain the origin of problem, provide *two solutions*, and discuss generalisations (sections "SAA as a guide to solve the problem of the initial example" and following). We especially point out what property renders the *t*-test valid but cross-validated decoding invalid and that employing SAA to potential solutions can help to test if they work as expected.

Note that the confounding effect that we demonstrate in this example is not a purely theoretical construct. It is also not simply remedied by increasing the number of subjects or the number of runs per subject; only substantially increasing the number of data points per run would help. We demonstrate this on real empirical data with multiple subjects in the supplement (SI2). A normal-sized empirical data set is also used in our second demonstration of another confounding effect below.

## The Same Analysis Approach

Problems like those in the initial example are hard to detect, because seemingly the experiment was designed correctly by using counterbalancing to neutralize a common confound. While an in-depth examination of design, analysis, and statistics might have alerted the researcher to the problem, it is often hard to determine what exactly to look for, especially for novel analysis methods with little practical experience. Performing empirical control analyses is a good idea, but can systematically fail if analysis methods differ between main and control analyses, as we demonstrated above.

In addition to theoretical examination and standard control analyses, we propose to perform the following types of analysis:

*Type 1) Apply the same analysis method used for experimental data to variables of the experimental design.* Perform positive and negative control analyses on synthetic noise-free data sets, each created from single variables of the experimental design, and analyse them with the main analysis method. Positive control analyses (see e.g. Fedoroff and Richardson, 2001) test if design variables that *should* influence the experimental outcome – typically the experimental variable – indeed yield significant results; failing these tests demonstrates that the experimental setup (design, analysis method, or their combination) are not suitable to detect the effect of interest. In more complex designs these should also test latent design variables that describe dependencies within the design. Negative control analyses test if design variables that *should not* influence the experimental outcome indeed do not yield significant results; failing these tests indicate that the variable is a potential confound, and/or that confound control did not work.

*Type 2) Apply the same analysis method used for experimental data to empirical control data.* The main analysis method is applied to additionally measured variables (e.g. reaction times, age, IQ). Since control data provide a proxy for the main results, here a result indicates how the main analysis would respond if the actual data were influenced by a confound.

*Type 3) Apply the same analysis method used for experimental data to synthetic null data.* Applying the same analysis to multiple realisations of synthetic null data tests if the false positive rate (the alpha level) is indeed as expected, and provides other general information on the null distribution of to be expected outcomes, such as range and shape.

We suggest to start SAA analysis as simple as possible For example, simple one-factor tests are very efficient at detecting confounds: They are easy to set up, have high diagnostic power, and we have found them to be useful in practice. After all, the aim of this article is to provide a framework to efficient-ly detect, avoid, and eliminate confounds, not to create unnecessary workload for experimenters.

**SAA to detect the problem of the initial example**

To illustrate SAA, we return to the initial example: An experiment with four runs and two trials per run, one for each experimental condition **A** and **B**, with presentation order counterbalanced across runs. Neurophysiological data is measured from a single voxel (ROI 1) and a larger region of interest (ROI 2). Additionally, reaction times are measured (Figure 3a). As assumed above, the neurophysiological example data are not influenced by the experimental condition, but only by the confounding factor "presentation order". In applying leave-one-run-out cross-validated classification we find 0% accuracy in both ROIs in the main analysis, leaving us with the question how to interpret this result (Figure 3b).

We now use SAA type 1 (Figure 3c): the same analysis on design variables. This experiment has the three design variables "experimental condition" (positive control), "run number" and "trial number" (both negative controls). The experimental condition can be translated into pseudo-data by using the assignment **A** = 1 and **B** = 2. The analysis result of 100% accuracy confirms that cross-validated classification could detect an effect of the experimental manipulation if there was one ("positive control analysis"). Run number (1–4) and trial number (1 or 2) can be directly used as pseudo-data. Here the analysis on run number results in the expected chance level of 50%, but the analysis on trial number results in 0% correct, providing a strong indication that a "trial order" confound could explain the observed below-chance accuracy (failed "negative control analysis"). Apparently, cross-validated classification is susceptible to this confound even though counterbalancing has been employed to counteract it.

Following SAA type 2, we next apply the same analysis to the reaction times as control data, and find again an accuracy of 0%. This provides evidence that the "trial order" confound indeed influences the data.

Finally, we employ SAA type 3 and apply the same analysis to simulated, random null data that contain neither an experimental effect nor a confound. Here the result will be different for each simulation, but performing many simulations we observe that the classification accuracy fluctuates around 50% on average, which is the expected chance level. This can be repeated with random data of different dimensionality and different distributions. See supplemental Table SI 1 (in section SI 3) for more details.

**a) Design variables and recorded data**

| Exp. variable/Cond. of Interest (label) | A B | A B | B A | B A |
|---|---|---|---|---|
| Other design variable (trial nr) | 1 2 | 1 2 | 1 2 | 1 2 |
| Neuro (ROI 1, 1d) | ☺☺ | ☺☺ | ☺☺ | ☺☺ |
| Neuro (ROI 2, nd) | ☺☺ | ☺☺ | ☺☺ | ☺☺ |
| Reaction Time [ms] | 100 200 | 100 200 | 100 200 | 100 200 |

**b) Main analysis**

| Exp. variable/Cond. of Interest (label) | A B | A B | B A | B A | Outc. | |
|---|---|---|---|---|---|---|
| ROI 1 data (1d) | 10 20 | 10 20 | 10 20 | 10 20 | 0% | ?? |
| ROI 2 data (nd) | [17 24; 8 21] | [17 24; 8 21] | [17 24; 8 21] | [17 24; 8 21] | 0% | ?? |

**c) Parallel SAA test cases**

| | | | | | Outc. | Exp. | As exp. |
|---|---|---|---|---|---|---|---|
| Cond. of Int. (recoded A:1, B:2) | 1 2 | 1 2 | 2 1 | 2 1 | 100% | 100% | ✓ |
| Other design variables (run nr) | 1 1 | 2 2 | 3 3 | 4 4 | 50% | 50% | ✓ |
| (trial nr) | 1 2 | 1 2 | 1 2 | 1 2 | 0% | 50% | ✗ |
| Reaction Time [ms] | 100 200 | 100 200 | 100 200 | 100 200 | 0% | ~50% | ✗ |
| Synthetic null data (1d or more) | 5 -3 | -4 1 | 2 4 | -3 -3 | ~50% | ~50% | ✓ |

**Figure 3.** The Same Analysis Approach (SAA) applied to the initial example. a) Design variables of initial example (experimental conditions/trial number) and assumed data (neural data from a one-dimensional and another high-dimensional ROI, reaction times). b) Features of neural data used in the main analysis. One data point (ROI 1: one-d, ROI 2: high-d) is available per trial. c) Parallel SAA analysis on test data: one data point is available per trial, either generated from design properties (condition of interest, run nr, trial nr), control data (RT), or synthetic null data. Abbreviations in figure: "Outc.": Outcome; "Exp.": Expected; "~50%": 50% plus/minus statistical deviation.

Had the experimental design contained additional variables, we could have systematically gone through all design variables, each time used the design variable instead of measured data, performed the same analysis on these values, and checked if the influence of this variable is as expected.

Note that, except for type 2, SAA does not rely on real data. Therefore, the problem with this combination of experimental design and analysis method could have actually been detected (and solved, see below) before data were collected.

**SAA as a guide to solve the problem of the initial example**

The analyses above demonstrate that the presentation order confound lead to below-chance classification and thereby explains the result of the main analysis. Further theoretical examination based on these results along the lines of Figure 1d reveals that the culprit is a mismatch between counterbalanced design and cross-validated analysis, in particular that the design factor "trial order" is *not counterbalanced within each training* and test *set*. One element that might be confusing in this context is the terminology: The analysis is actually "balanced" in the sense in which the term is normally used in cross-validation, which is to say that the number of training samples per class are equal in each partition. Having both "balanced" data (presentation order) as well as a "balanced" cross-validation scheme (number of sample per class) makes it difficult to detect that the cross-validation is both "balanced and "not balanced" at the same time.

Note that the origin of the problem in this specific example is indeed only the missing counterbalancing in each cross-validation fold, and neither the analysis type (*t*-test vs decoding) nor the dimensionality of the data (the demonstration actually is one-dimensional). Cross-validated MANOVA (Allefeld and Haynes, 2014), cross-validated Mahalanobis distance (Diedrichsen et al., 2016) used in RSA (Kriegeskorte et al., 2008), or any other cross-validated distance measure will all suffer from the same problem and systematically estimate negative distances, which are as confusing as below-chance results.

**Two potential solutions and SAA to verify whether they work**

One possible remedy for this problem is not to use counterbalancing but randomization, i.e. to randomly decide for each run independently whether to use the trial order **AB** or **BA**. We can now employ SAA again to *test if the solution indeed works as expected*, by re-running the same analysis on the design variable "trial number" for randomized designs. When simulating many experiments, we find that the average classification accuracy is indeed 50% (the chance level), i.e. that the confound is statistically controlled. Looking at the individual outcomes, however, we find that 50% accuracy itself never occurs; rather, 0% occurs in 3/8, 75% in 1/2, and 100% in 1/8 of all randomizations (supplemental Figure A). SAA thus revealed that randomization does not seem an ideal solution in this context.

Another possibility to solve the problem would be to keep the design, but to use a validation scheme which ensures that the confound is counterbalanced in each test set, i.e. that each contains equally many **AB** and **BA** runs. This can be achieved by leaving out two runs (training sets in the four folds: runs 1 & 2, runs 3 & 4, runs 1 & 4, runs 2 & 3; supplemental Figure B). We can again employ SAA to *test if this new analysis solves the problem*. This time the result is indeed 50% for every single experiment, and not just on average as above.

These two possibilities are of course not exhaustive. Since in this example the problem is related to the way cross-validation is implemented, another alternative would be to replace classification accuracy by a (multivariate) test statistic that does not need cross-validation.

Please note that the example here has been deliberately chosen to be as small as possible. The demonstrated systematic negative bias will, however, also occur in larger, real datasets if trial order has an effect on the data and leave-one-run-out cross-validation is used. The negative bias may not be as extreme as in the example, but can easily be large enough to suppress real effects and/or lead to confusing significant below-chance accuracies. See supplemental section SI 2 for a demonstration on a real empirical dataset.

**Related work and generalisation (initial example)**

Three other causes for systematic below-chance results have been described previously. The first has been provided by Kohavi (1995), who noted that a majority classifier (that simply predicts for each test data the label that is most common in the training set ignoring any properties of the data) will yield 0% when leave-one-out cross-validation is employed on a balanced data set (with equally many samples per class). While the example is simpler than ours and critically depends on different numbers of exemplars per class, it already has the same general structure as ours, because again balancing is ignored when splitting data into training and test sets. The second example is "anti-learning" (Kowalczyk, 2007), which demonstrates that datasets with specific properties will always yield below-chance accuracies for a large number of classifiers, independent of any specific design property or validation scheme. The third cause hinges on using the binomial test for single cross-validated accuracy estimates, which will yield too many significant below-chance results (Jamalabadi et al., 2016; Görgen et al., 2014) and above chance results (Noirhomme et al., 2014; Görgen et al., 2014). Another scenario in which counterbalancing also unexpectedly fails to control a confounding factor in MVPA has been described by Todd et al. (2013). It differs from our example because in theirs individual decoding analyses are calculated for each unit (subjects in their example, runs in ours), whereas only a single decoding analysis using all units is calculated in our example. Other major differences are that it causes above chance results, not below chance results, and that it does not depend on any particular cross-validation scheme, which is the crux in our example.

Our example demonstrates that systematic below-chance classification accuracies can be caused by a *design–analysis mismatch*, which can even occur when employing only basic experimental methodology. In the specific example above, the design variable "trial order" was controlled. The problem, however, is not specific to controlling time or sequence effects; the same logic applies to counterbalancing any other variable against the experimental variable. In general, it often has unexpected consequences if design features which are implemented with respect to the full data set are ignored when data is split into training and test sets for cross-validation. Examples for this are cases where each class has an equal number of samples in the full data set but differing numbers in each training and test set, or cases of "dissolving strata" such as the assignment of patients and their matched controls to different partitions.

## Principles for setting up SAA

In this section, we give a non-exhaustive overview over possible forms of SAA and the different aspects that have to be considered in setting up an analysis. Supplement section SI 4 provides more in-depth explanations of components of individual test cases, and section SI 6 demonstrates the necessary steps to perform SAA for the concrete empirical example below.

**Test data**

*Design variables:* These can be explicit design variables such as the experimental condition or the level of a factor in a factorial design, or implicit design variables such as the sequential number of the trial within run or the repetition number of a stimulus.

*Control data:* These are additionally recorded data such as reaction times, error rates, motion correction parameters, eye-tracking data etc. Possible across-subject data include age, gender, IQ, or personality scores.

*Simulated data:* Simulations open a wide range of possibilities. Data may be generated so that there is no effect (null data) or there is a specific effect, that a confound is present or not present, or combinations thereof. They may be simplistic, for example data consisting of only 1s (constant data), or they may come from a generative model attempting to capture as many aspects of real data as possible (distribution, autocorrelation across time and space, hemodynamic response, effect size, variation across measurements, trials, runs, and subjects). A special case are modified data from the same experiment, e.g. shifted by one trial (Soon et al., 2014), or experimental data unrelated to the experiment, such as resting-state data (Eklund et al., 2016).

*Mapping function:* In some cases, test data may be in a form that cannot be processed by the "same analysis". An example is the experimental condition, which is a nominal label and therefore not compatible with a classifier that expects numerical input. Such categorical data may be mapped to input data in several ways: Conditions are arbitrarily assigned numerical values (see example above), or encoded as multiple dummy variables (1 if a trial belongs to a condition, 0 otherwise), or assigned to randomly chosen multivariate patterns. Another case are analyses that use intrinsically multivariate measures such as pattern correlation or cosine distance, e.g. in representational similarity analyses (RSA; Kriegeskorte et al., 2008). Here, simple mathematical or statistical models can be used to create multivariate data, where similarities are determined by the input variable. Indeed, there is high value in creating different test cases that all map the same variable to test data, but with different mapping functions, to understand how the analysis pipeline reacts to input that might be encoded different than expected (e.g. if it is not clear which coding scheme the brain employs to encode a specific stimulus). Depending on the complexity of the mapping function, there is a continuum between SAA on a simple design variable and a full-blown simulation.

**Test range**

The Same Analysis Approach can be applied to different analysis ranges: A complete pipeline, single parts, or specific combinations of parts. In an MVPA study, these parts may be pre-processing of data, extraction of single-trial or run-wise values, cross-validated classification, second-level analysis, and statistical inference. Depending on the range, the form of both test data and inspected outcomes changes, e.g. time series, trial-wise values, run-wise values, accuracies, test statistic values, *p*-values, or statistical significance.

**Test case**

Together, each combination of test data, mapping function, test range, and outcome specifies a unique *test case*.

**Expected outcome**

Whenever possible, each test case should come with a defined expectation (e.g. chance level classification if there is no effect), and interpretations if the expectation is fulfilled or violated. Depending on the test data (see below), an expectation may be a specific value (e.g. an accuracy of 50%) or a distributional property (e.g. average accuracy 50%).

**Deterministic vs stochastic tests**

When the test data are fixed, e.g. noise-free pseudo-data generated by a deterministic mapping from a design variable, there is only one corresponding analysis result, and the interpretation of the result depends on this single fixed value. For noisy data like experimental control data the outcome is still fixed, but its interpretation is not straightforward and a statistical test may be necessary to determine whether the result is significant. In a simulation incorporating random variation, the simulation has to be run a sufficient number of times to assess properties of the distribution of outcome values, e.g. mean, variance, or number of significant outcomes. For the latter, statistical testing and simulations can be combined by looking e.g. at the frequency with which the statistical test indicates a significant result across simulation runs, to determine whether the test is valid under the given circumstances.

**Recommendations when using many statistical tests**

It is simple to implement a large number of SAA tests, especially using simulated data. If the results are assessed by a statistical test, the number of false positives will increase with the number of tests, so that the significance level has to be adjusted. This raises the question how to balance between sensitivity and specificity for possible confounds, and how to efficiently detect problems within many test outcomes. For this purpose, we suggest the following measures:

*Adjusting the significance level only for less important tests.* Tests should be separated into a small number of important tests, which are targeted at potential confounds that are expected to exert a strong influence, and a possibly large number of less important tests that are only performed to be on the safe side. For the first class, sensitivity (as controlled by the significance level) is kept high, while the second class is corrected for multiple comparisons.

*A priori checks vs problem diagnosis.* When SAA is set up prior to data collection (see below) or when no signs of a problem exist in the analysis of experimental data, the sensitivity can be lower than when trying to find the source of a concrete problem which is evident in main analysis.

*Sorting test cases by influence.* Tests should be sorted according to whether a violation in one test is likely to imply a violation in another test, because the problems targeted by each overlap. For example, if a test on null data shows an unexpected result, it is likely that there is a very deep-seated and general problem which also influences the outcomes of other, more specific tests.

*Interpretation of results.* Problem diagnosis should not rest simply on whether a statistical test gives a significant result, but the researcher should use their judgement to decide whether a confound is likely to be relevant in the main analysis. More realistic simulations can help to assess the practical impact of a confound.

**Correlating SAA outcomes and main outcomes as additional check or to detect location-specificity**

If multiple SAA test cases indicate potential confounds, only some of them may actually affect the main analysis. To check this, correlations can be calculated between the outcomes of one or multiple SAA test cases and the outcomes of the main analysis. As with any statistical test, a negative result does not mean that the tested variable is not a potential confound, but a positive result strongly indicates that it is (see Reverberi et al., 2012 for an application example). Moreover, if the same main analysis is performed on different segments of the data, e.g. brain regions or time points, correlations to SAA outcomes can be calculated for each segment to detect location-specific confounding effects, e.g. a confound may only affect motor cortex but not visual cortex.

**When to use SAA**

SAA helps to find solutions when experimental data have already been acquired and their analysis indicates that there may be a problem; in some cases, however, it may come too late at this phase. We therefore recommend to use SAA systematically during different phases of a study (Figure 4):

*Design phase.* Tests can already be set up when designing and implementing an experiment to ensure that the analysis pipeline works as expected and that the design matches the analysis.

*Piloting phase.* During behavioural pre-tests or pilot studies tests can be used to check whether potential confounds are present in participants' responses.

*Main analysis phase.* After data collection, tests can be run on control data to check whether corresponding confounds may be present, or in the worst case, to diagnose a problem that has become apparent in the main data analysis.

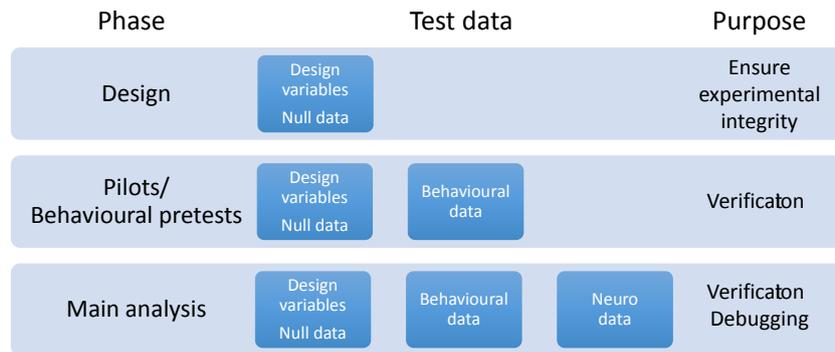

**Figure 4.** Guideline for using SAA in different phases of a study.

## Empirical Example: Variance confound in classification

In this section, we demonstrate how to use SAA to diagnose a problem on real empirical data. A researcher performs an experiment where participants press a button with either the left or right index finger in response to visual stimuli. Left button presses are more frequent than right button presses, 12 vs 3 trials per run (following e.g. an oddball paradigm, Squires et al., 1975). BOLD data are recorded in 6 runs from 17 participants. To identify brain regions that carry information about which button was pressed, the researcher applies leave-one-run-out cross-validated classification to parameter estimates from voxels within a searchlight, using a linear support vector machine. For a time-resolved analysis, they use finite impulse response (FIR) regressors comprising 16 two-second time bins (cf. Kriegeskorte et al., 2006; Soon et al., 2008). Because they are aware that imbalanced data pose a problem for many classification algorithms (He and Garcia, 2009), they use a single set of regressors for modelling left and right button presses, respectively (Allefeld and Haynes, 2014; Haxby et al., 2011; Norman et al., 2006). For each FIR time bin, this yields a single parameter estimate image per condition and run, all of which are then used for time-resolved searchlight classification. Subject-wise classification accuracy maps are then entered into a second-level *t*-test across subjects against the chance level of 50%.[2]

There are clear expectations for the result of this analysis. First, information should be localized mainly in motor regions because the analysis contrasts two different movement conditions. Second, above-chance classification should be possible no earlier than 4s after button press because of the hemodynamic delay. The results, however, show significant information in large regions across the entire brain, and already at 0-2s after button press (Figure 5, top). Apparently, something in the analysis went wrong.

---

[2] This example was constructed using data from an unpublished study on rule representation. Preprocessing, parameter estimation, and second-level analysis of fMRI data were performed with SPM8 (Wellcome Department of Imaging Neuroscience) and searchlight classification with The Decoding Toolbox (Hebart et al., 2015) using LIBSVM (Chang and Lin, 2011).

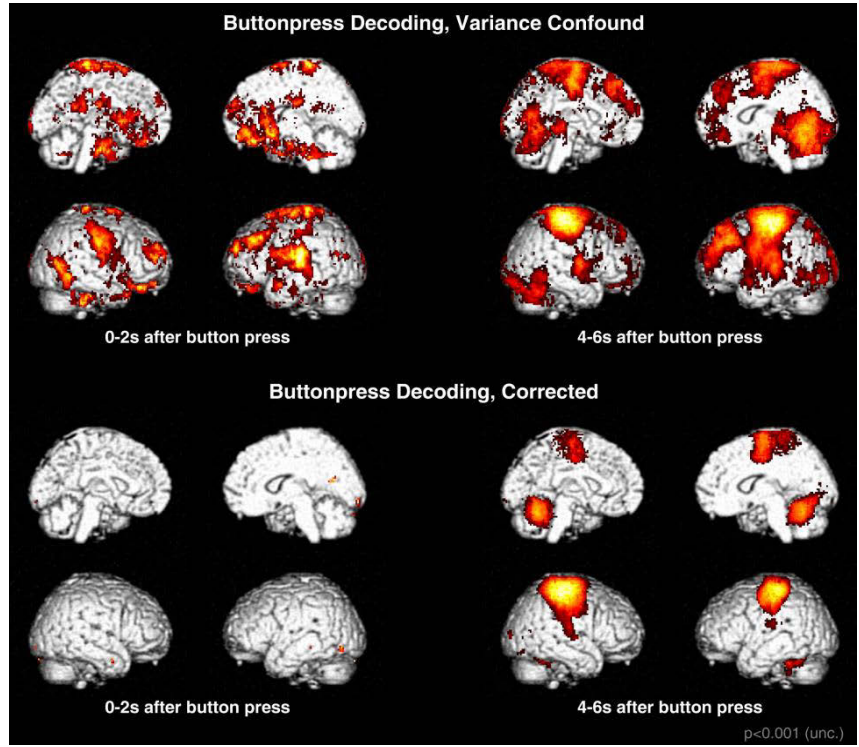

**Figure 5.** Results of confounded and corrected example fMRI analyses. Top: Significant results of button press classification with variance confound on real data 0–2s and 4–6s after button press. Run-wise GLM parameter estimates were calculated using 12 trials for left and 3 trials for right button presses. Bottom: Same as above, but using 12 left and 12 right button presses to calculate run-wise estimates. – All displayed voxels show significant effects at $p \leq 0.001$ uncorrected. All larger clusters are also significant at $p \leq 0.05$ FWEc-corrected; only in the corrected analysis at 0-2s no cluster survives FWEc correction (bottom left). Supplement Figures D (SI 5) contain more combinations & time bins.

**SAA setup**

The researcher wants to use SAA to diagnose the suspected problem, checking for temporal, attention, and sequence effects, as well as details of the task. They create test cases by making the following decisions:

*Test data:*

Synthetic noise-free positive test
– The condition of interest itself ("side")

Empirical negative tests
– Attention effects: response time, correctness of response

Synthetic noise-free negative tests
– Temporal effects: number of trials ("ntrial"), time of button press, target onset
– Sequence effects: value of all these variables from the previous trial ("t-1")
– Additional: Constant data that has the value 1 for each trial ("const")

Synthetic null-distribution negative tests
- 10,000 one-dimensional random null datasets ("randn1", "randn2", etc.) drawn from the standard normal distribution)

Numerical test data (e.g. time, number of button presses) are used as input values for the analysis as-is (e.g. the values 1, 2, …, for trial numbers). Categorical data (side of button press) are mapped to dummy variables (here a two-dimensional vector, that is [1 0] for trials that expect a left button press, and [0 1] for trials that expect a right button press).

*Test range:* Test data are generated on the level of single-trial values, and the whole analysis from there to the second-level *t*-test on accuracies is considered. The analysis steps in this range are: 1) computing run-wise parameter estimates, 2) leave-one-run-out cross-validated classification, and 3) a group level *t*-test applied to subject-wise classification accuracies. Outcomes are subject-wise accuracies (visualized through box plots), *p*-values of the second level *t*-test, and the frequency with which the test indicates significance for null data.

To increase the sensitivity of SAA, the researcher sorts the test cases into different categories, labelled "sanity checks", "design random" (test cases for which the result can vary for different test cases), and "control data" (Figure 6). Supplement section SI 6 provides a more detailed explanation including the concrete steps to setup this SAA analysis.

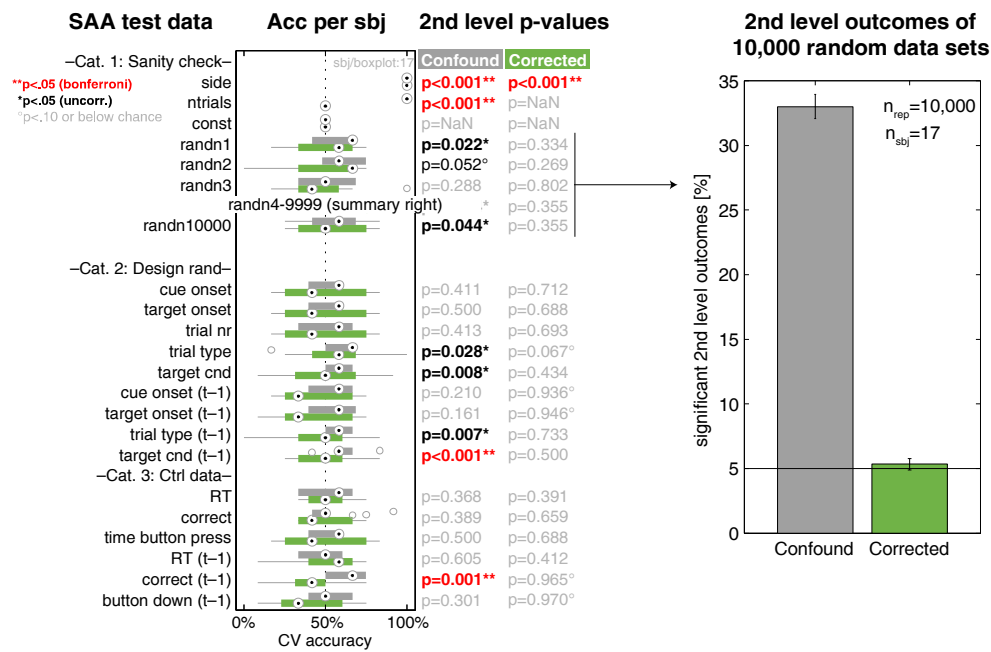

**Figure 6.** SAA results for different test data. Left panel: Distribution of accuracies per subjects as box plots with medians (filled circles) and outliers (empty circles) before (grey) and after correction (green). Small dotted line marks the chance level of 50%. "2nd level *p*-values" column provides the *p*-values for a one-sided *t*-test across subjects against 50%. Right panel: Summary for the 10,000 simulated random null data sets ("randn1" – "randn10000"), showing the relative frequency of cases where the SAA result was significant ($p \leq 0.05$).

**Interpretation of SAA results**

The results shown in Figure 6 show significant effects for several negative control analyses, for which no effect was expected.

Focusing on the "sanity check" category first, the researcher is reassured by the outcome of the positive control analysis "side", which confirms that the analysis is able to distinguish left and right button presses if there is a difference between the corresponding trials. There is also a significant effect for "ntrial", the number of trials per condition, which is not surprising since there is a systematic difference between conditions in the number of trials (3 vs 12). By contrast, there is an unexpectedly high number of significant results for random null data: the second-level $t$-test rejected the null hypothesis in 33% ($CI_{95\%}$=[32.1%, 33.9%]) of the 10,000 instances (at $\alpha$ = 0.05) instead of the expected 5%.

The SAA has therefore confirmed the suspicion that there is a problem. The increased false positive rate of the null simulations strongly suggests that it has to be a more general aspect of the design or some property of the analysis procedure, because neither the experimental variable nor any other design factor had any influence on the simulated null data. A peculiar property of the design is the different number of trials in the two conditions. The researcher assumed to have dealt with this by applying classification not to single-trial data but to run-wise parameter estimates – but what if this was not enough?

The researcher checks this hypothesis by modifying the SAA analysis so that equally many trials are used to calculate the estimates for left and right button presses in each run. Indeed, after this correction (green elements in Figure 6), the number of "significant" results in the 10,000 instances of null data drops to 5.4% ($CI_{95\%}$=[4.9%, 5.8%]), consistent with a false positive rate of 0.05. The only test case that remains significant is the positive control using the variable of interest itself ("side"), which is how it should be.

The result of the corrected analysis on the fMRI data (Figure 5, bottom) confirms that the apparent confound has been removed; there is no significant effect present in the 0–2s time bin, and effects in the 4–6s time bin are located in motor and sensory regions as expected. Supplemental Figures D.1-D.8 in supplement section SI 5 show the time-resolved results for all combinations of 3, 6, and 12 button presses from each side, illustrating that problem is indeed caused by the imbalance between trials and not by e.g. differences in power; choosing the same number of left and right trials always solves the problem.

**Cause of the problem: Variance-based linear classification**

As described above, SAA can help to diagnose a problem and to quickly check whether an approach to resolve it is likely to work. It does not by itself reveal its cause – this is left to the researcher. To conclude this example, we briefly explain how the problem arose.

During experimental setup, the reasoning of the researcher was the following: 1) Classifiers are known to be sensitive to imbalanced training data, therefore classification is applied to run-wise parameter estimates, which are essentially averages across trials. 2) Linear classifiers are sensitive to linear differences between class-specific data distributions, i.e. differences in the class means. 3) If there is no effect, trial-wise data from both classes comes from the same distribution, and averaging over more or fewer trials does not change the mean.

The mistake in this argument is that while the difference in number of trials per condition does not change the mean, it does change the variance of run-wise estimates. And contrary to common assumption (Kamitani and Tong, 2005; Naselaris et al., 2011; Norman et al., 2006), linear classifiers can not only use differences in mean, but also differences in variance to achieve above-chance classification (Figure 7). This behaviour is not limited to specific types of linear classifiers; it applies even to classifiers utilizing the means (centroids) of the data, such as nearest centroid classifiers or linear discriminant analysis (explanation in caption of Figure 7, especially panels c,d). More generally, linear classification based on the variance of parameter estimates can come about by differences in the estimability of regressors (Hebart and Baker, this issue).

Note that successful linear classification that is based on differences in variance is not a confound, because the classifier reveals a difference that truthfully exists in the data (see also Davis et al., 2014 for other effects of variability in MVPA). It can, however, be an *interpretation error* if this is interpreted as showing a linear (mean) difference between conditions. In contrast, the empirical example contains a true confound, because the data from both classes do not come from different distributions, but the variance difference between both is *induced* during the analysis (Figure 7a,b). A detailed simulation for SVMs and nearest centroid classifiers can be found in the supplemental section SI 7 (Figures E.1 and E.2).

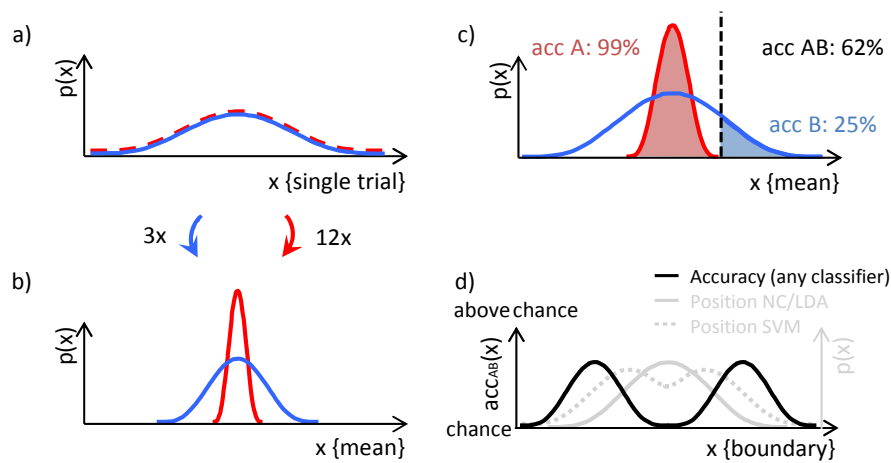

**Figure 7.** Induction of variance difference by design and successful variance classification with a linear classifier. a) Original probability distribution (one-dimensional) of single trial values for two classes (blue, red). b) Averaging (or regressing) different

numbers of trials creates distributions that have the same mean but different variance. c) Example of a linear classifier (classification boundary: black dashed line) that classifies between both classes above chance using the nonlinear variance difference between the two. d) Expected accuracy for classifiers with a decision boundary at different positions (black line), and probability distribution where a nearest centroid classifier or linear discriminant analysis (NC/LDA; grey solid line) or an SVM (grey dashed line) place the boundaries. The expected accuracy (black line) will at minimum be at chance level (when placed exactly at the common mean of both distributions, or at plus or minus infinity) and otherwise above chance. Because the position of the boundary varies (grey lines), the expected accuracy for classifying between classes that differ in the nonlinear mean using a linear classifier is above chance. Note that successful linear classification between data classes that only differ in variance (panels c,d) is not a confound, because the classifier truthfully reveals a difference that exists in the data. It can however be an interpretation error if this is interpreted as showing a linear (mean) difference. In contrast, the confound in the example arises because the variance between both classes is induced during the analysis (panels a,b). Detailed simulations can be found in supplemental section SI 7.

**Related work and generalisation (empirical example)**

The main aim of this example was to demonstrate how SAA can be employed in practice. However, the example is also interesting in itself. Averaging before classification and feature extraction from multiple trials before more complex data analyses methods are standard analysis procedures. We advise to test for potential confounding effects through null simulations here. Another important point relates to the inference from classification analyses: Since linear classifiers can successfully extract nonlinear information, successful linear classification does not allow direct inference on the linear versus nonlinear nature of representations (e.g. Kamitani and Tong, 2005; Norman et al., 2006; Naselaris et al., 2011; Diedrichsen and Kriegeskorte, 2017; Friston, 2009).

The example also demonstrates that confounds can arise through a combination of analysis steps that pose no problems individually. These can be detected by simple simulations on synthetic null data if the same analysis is employed.

## Discussion

In this paper, we advocated to systematically check experimental design, data, analysis methods, and statistical inference, in order to cope with the challenges and possible pitfalls of novel methods in neuroimaging (including MVPA). These methods sometimes fail to conform to researchers' expectations and intuitions. This leads (a) to situations in which confounding influences are not controlled and consequently spurious effects are observed or true effects fail to be identified, or (b) to overly optimistic or pessimistic effect size estimates. We propose to not blindly rely on such expectations and intui-tions but to explicitly check them, by applying the *same analysis* used for experimental data also to design variables, control data, and simulated data. We now discuss a number of points that may need further clarification.

**Keep it simple.** A main focus of this paper is to introduce design principles to create *efficient* control tests. We made a number of suggestions in this paper how this can be achieved. One main suggestion is "keep it simple"; other suggestions are to perform positive and negative control analyses on many simple control datasets that are each influenced by only one synthetic or empirical variable, and to create "time-shifted" datasets by using variable values from the previous trial to detect sequence effects which are common confounds in neuroimaging. Further recommendations to keep SAA effective include adaptive alarm rate thresholds and correlating SAA to main analysis outcomes. However, we do not believe that these are the ultimate and only principles to set up efficient tests, nor that they fit all experimental paradigms. We rather conceive them as first suggestions, and hope that further principles for efficient tests will emerge from employing SAA in practice.

**When to employ SAA.** SAA can be used to diagnose problems that have already become apparent, but we recommend to use it continuously through all phases of a study – planning, piloting, and final analysis – to become aware of possible problems as early as possible. Side benefits of this practice are that it encourages to consider details of the analysis already at the design phase and therefore to tailor the design to the questions one wants to ask; that it can be used for power analysis (if sufficiently realistic simulations are implemented); and that it helps to detect simple programming errors (both in design and analysis, because SAA tests depend on both). Indeed, the sole process of setting up SAA at the design phase can prevent programming errors in the first place, because the coding scheme of variable names and content are fresh in mind when programming design and analysis at the same time, reducing the risk of confusion between both. In contrast, if time passes between setting up design and analysis, e.g. when data is recorded, chances to confuse variable names or coding schemes are much higher. SAA might also facilitate design optimization, but we believe that further investigation of potential negative side effects is necessary.

**About our examples.** In addition to describing and detailing SAA in general, we illustrated it in two concrete examples. Their main function in this paper is to spell out in detail how SAA can be applied and how it uncovers potential problems with a given data set or design.

However, both of them are also relevant on their own, because they reveal two relevant problems in MVPA: The initial example demonstrates that the classic strategy of counterbalancing the experimental design (leading to what is also known as a crossover designs) to control a confound can become ineffectual if combined with an analysis method that uses cross-validation[3]. The empirical example shows that differences only in variances yield successful linear classification, specifically demonstrating that inferring linear differences from linear classification would be invalid (see "Cause of the problem: Variance-based linear classification"). In addition, we demonstrate that analyses of control data can fail, even if to-be-controlled effects are present, when different methods are employed for control and experimental analyses. The fact that neither example depends on the dimensionality of the data (both work for multi- and univariate data) demonstrates that unexpected confounds are also not specifically bound to multivariate analysis, but can occur for univariate analyses as well.

---

[3] In the specific example, the design variable "trial order" was controlled, but the same logic applies to any other counterbalanced variable.

**Relevance.** The fact that SAA would detect our examples as well as examples from the recent literature, both MVPA specific (Todd et al., 2013; Woolgar et al., 2014; Noirhomme et al., 2014; Görgen et al., 2014; Jamalabadi et al., 2016) and more general (Kriegeskorte et al., 2009; Vul et al., 2009; Mumford et al., 2015), demonstrates the potential of SAA in aiding to detect easy-to-overlook problems. We have found it helpful in personal work, and are looking forward to seeing whether or not that will be the case in general. Finally, we see employing the same analysis method as especially important for control analyses, at least in addition to other analysis methods, because – as demonstrated – they can fail their purposes when different analysis methods are employed.

**Not too few data; more data no remedy.**[4] A common misconception is that confounding effects occur only for small data sets, and that more data would reduce those confounds. While more data can help reducing effects of non-systematic confounds, simply adding more data is no universal solution, specifically when confounding effects are systematically induced by design, such as in the examples that we demonstrate here. Indeed, the empirical example already has a normal-sized sample, and because the confounding effect is present in each subject, more subjects would even increase the effect strength. The same holds for the initial example if the presented design would be used for multiple subjects and a test would be applied on the group level (see supplemental section SI2). It would also stay a potential confound if the number of runs is increased. In the idealized case with no noise and no effect (as in the example), the classification accuracy would stay at 0% correct. In real data, for increasing number of runs the importance of the confound depends more and more on the relative effect sizes of confound and experimental effect. If no experimental effect is present, the primary effect measure (classification accuracy) will come closer to chance level, but because the null distribution becomes narrower, the confounding effect could still have a significant impact.

**Differences between SAA and simulation studies.** SAA shares aspects with standard simulation studies that are routinely used to demonstrate merits and pitfalls of particular design or analysis methods. Like SAA, simulation studies demonstrate their claims through computation.

---

[4] Also known as the "more data no cry" fallacy.

SAA however differs from simulation studies in important aspects. First, it avoids a particular problem in simulation studies, which is to choose which settings are important to demonstrate generality. Because SAA is used for a *particular* experiment, most parameters (such as number of subjects, etc.) are fixed. Second, simulation studies typically include complex realistic simulations, to demonstrate the operation of a method in realistic scenarios. In contrast, SAA is employed to perform sanity checks, which we believe can be effectively done with simple simulations. Whether or not this is the case, and which additional principles can help to create useful control analyses is an open question, that we believe will need to employ SAA in practice. Thus, SAA is not a theoretical tool to demonstrate a claim; it is an empirical tool to help creating better designs and analyses.

**SAA and unit testing.** SAA has been inspired by the practice of "unit testing" in software development (Myers et al., 2011), i.e. writing software in the form of modules that each can be tested independently (for internal function) and in combination (for adherence to interfaces between modules). The situation in software development is insofar similar to that in neuroimaging that in principle the validity of an algorithm may be strictly proven, but the multitude of newly produced code makes that practically impossible. In contrast to unit-testing, SAA however does not test software modules, e.g. functions of an analysis package, but instead design–analysis combinations of specific experiments.

**SAA in other fields.** SAA shares its rationale with a number of other scientific approaches. It follows the same logic as the routine use of positive and negative controls in disciplines like chemistry or molecular/cell biology (Fedoroff and Richardson, 2001; Johnson and Besselsen, 2002), where the working of the full analysis pipeline is tested for every experimental data again by analysing positive and negative probes alongside with the experimental data, e.g. during PCR, or in medicine, e.g. using diluent and histamine as controls in skin prick testing during allergy diagnosis (Rusznak and Davies, 1998).

**Not a general solution.** We would like to point out that although SAA is a tool that can be generally applied to data analysis pipelines and is not specialized to find specific kinds of problems in specific kinds of analyses, there is no guarantee that it will help to detect any kind of problem in any kind of analysis. Moreover, SAA in itself does not solve any problem, but merely points the researcher to possible problems that then have to be resolved on a case-by-case basis.

**Conclusion.** We hope that new developments in neuroimaging data analysis will in the long term lead to the establishment of a new corpus, and in particular that the heuristics of machine learning methods will be backed up by and integrated into the theory of statistical inference (Efron and Hastie, 2016). However, we believe that testing experiments with SAA provides a highly efficient additional safeguard to detect, avoid, and eliminate confounds, and can therefore help improving quality and replicability of experimental research.

## Acknowledgments

The authors declare no conflict of interest. This work was supported by the German Research Foundation (DFG Grant GRK1589/1 & FK:JA945/3-1). M.N.H. was supported by the German Ministry of Education and Research (BMBF, Grant No. 01GQ1006), by the Intramural Research Program of the National Institute of Mental Health (Protocol 93-M-380170, NCT00001360), and a Feodor-Lynen fellowship of the Humboldt Foundation.

# Supplemental Information to
# The Same Analysis Approach:
# Practical protection against the pitfalls of novel neuroimaging analysis methods

Kai Görgen[a], Martin N. Hebart[bc], Carsten Allefeld[a]*, John-Dylan Haynes[ade]*

[a] Charité – Universitätsmedizin Berlin, corporate member of Freie Universität Berlin, Humboldt-Universität zu Berlin, and Berlin Institute of Health (BIH); Bernstein Center for Computational Neuroscience, Berlin Center for Advanced Neuroimaging, Department of Neurology, and Excellence Cluster NeuroCure; 10117 Berlin, Germany
[b] Department of Systems Neuroscience, University Medical Center Hamburg-Eppendorf, Martinistr. 52, 20251 Hamburg, Germany
[c] Section on Learning and Plasticity, Laboratory of Brain & Cognition, National Institute of Mental Health, National Institutes of Health, Bethesda MD, USA
[d] Humboldt-Universität zu Berlin, Berlin School of Mind and Brain and Institute of Psychology; 10099 Berlin, Germany
[e] Technische Universität Dresden; SFB 940 Volition and Cognitive Control; 01069 Dresden, Germany
* These authors contributed equally to this work

**Content**

# 1  Possible solutions to the initial example

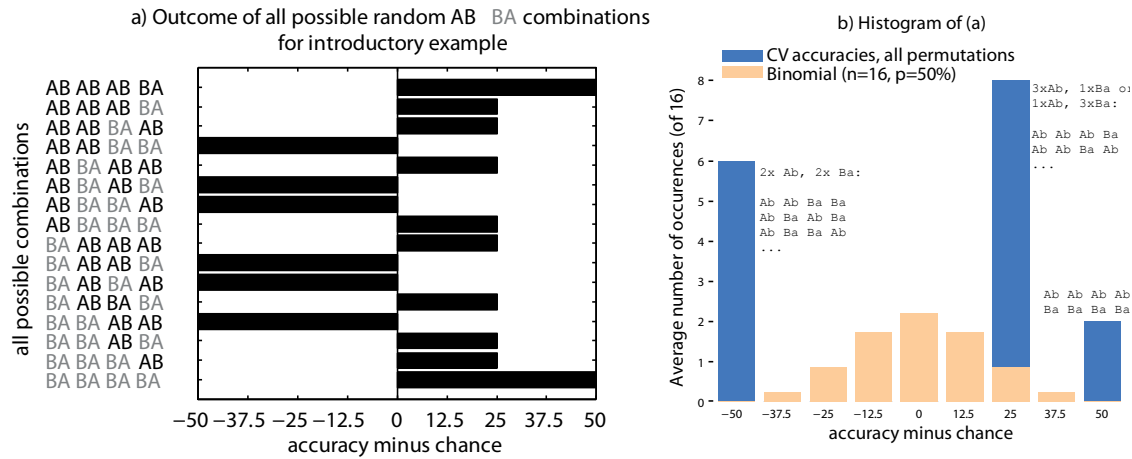

**Figure SI A. First possible solution to the initial problem: statistical control.** The first solution to "solve" the initial problem is through statistical control, i.e. to randomly select a presentation order for each run individually. a) Outcome of all potential experiments with four runs (AB: measuring A before B; BA: B before A). When the data exclusively depends on "trial order" as in the initial example, the expected outcome (the average across all possible realizations) is 50% correct and thus the cross-validated estimate is unbiased as expected. Each single experiment does however strongly deviate from 50%, and indeed no single experiment will reach 50% (but either 0%, 75%, or 100%; black bars). b) Data from panel a as histogram (blue) compared to the binomial distribution (orange).

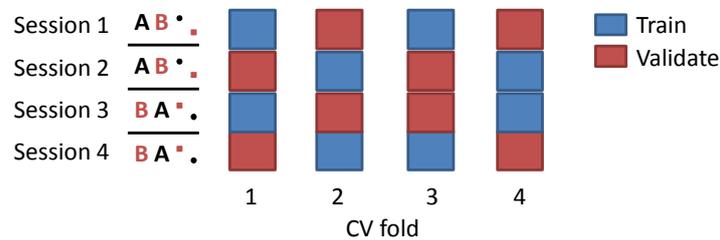

**Figure SI B. Second possible solution to the initial problem: counterbalance each training set.** Ensuring that the factor that should be counterbalanced (here the presentation order AB or BA) will indeed be counterbalanced in the *training* set of *each cross-validation fold*. This has the advantage that the outcome of each individual experiment will be unbiased.

## 2 Empirical demonstration of the initial example

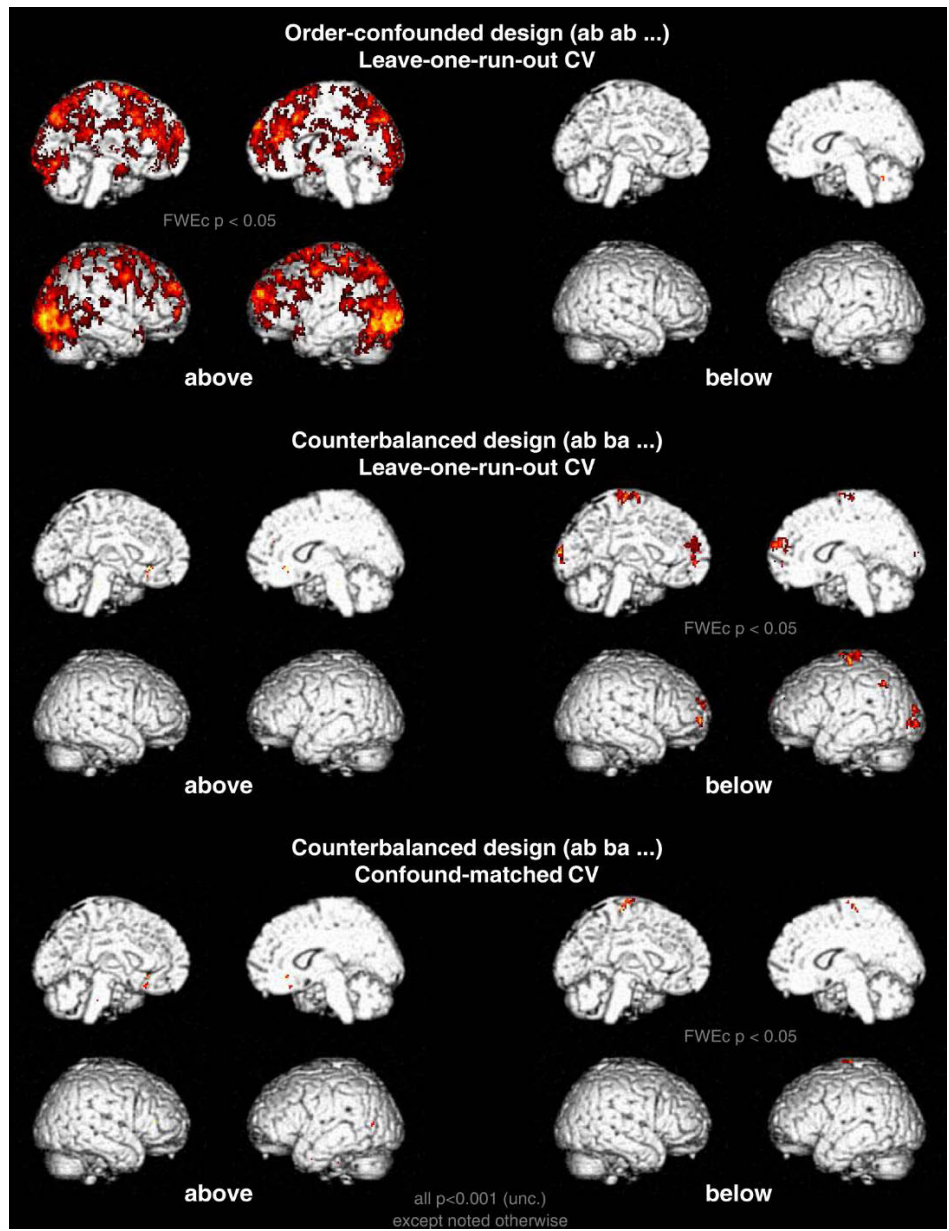

**Figure SI C. Demonstrating the effect of the initial example on a real dataset.** The same dataset as for the empirical example was used for the example (n=17 participants, 6 runs per participant; see empirical example in main paper for full setup). One HRF regressor was calculated for the first and the second trial for each run and participant. In general, searchlight classification analyses were then conducted for each participant, and the resulting accuracy maps were statistically analysed with a second level test (see main paper for more details). Specifically, three decoding analysis were conducted: **Analysis 1 [top])** classifying first vs second trials with leave-one-run-out; the results (top) demonstrate that a strong order confound exists in real fMRI data. **Analysis 2 [middle])** Again leave-one-run-out cross-validate classification, but this time counterbalancing first and second trials across runs, i.e. have three runs with the first trials in condition A and three with the second trials, and vice versa for condition B. The results (middle) demonstrate that outcome is as predicted in the initial example: While the order effect does not lead to false positives anymore, it does lead to significant below chance accuracies (that might be confusing). **Analysis 3 [bottom])** The same analysis classification as in analysis 2, but modifying the cross-validation such that in each training set the factor trial order is counterbalanced again, i.e. using all combinations of two AB and two BA runs to train and the remaining data to test. As predicted, the results (bottom) have neither false positive nor false below-chance accuracies. Thus, this analysis achieved what counterbalancing was expected to achieve in the first place: It neutralized the confounding effect of trial order, i.e. it did led to neither above nor below chance accuracy biases.

# 3   SAA for the introductory example

**Table SI 1 – SAA for the introductory example.** Abbreviated column names: Expectation (Exp.), result (Res.), as expected (as exp.). See SI section 4 below for explanation of column headers. ~ indicates that the values not fixed but stochastic, i.e. that single test outcomes vary around the provided value.

| Test case # | Intention | Diagnostic information | Mapping function | Mapped data | | | | Exp. | Res. | As exp. |
|---|---|---|---|---|---|---|---|---|---|---|
| | | | | A B | A B | B A | B A | | | |
| 1 | Test analysis can be significant | Exp. cond. | y=[A:1;B:2] | 1 2 | 1 2 | 2 1 | 2 1 | 100% | 100% | ok |
| 2 | Test factor trial nr. has no influence on result (esp. because it's counterbalanced) | Trial nr. | y=x | 1 2 | 1 2 | 2 1 | 2 1 | 50% | 0% | !!! |
| 3 | Test behavioural difference (influence of hidden factors) | RT | y=x | 100 200 | 100 200 | 200 100 | 200 100 | 50% | 0% | !!! |
| 4a | Test false positive rate on null data, 1D, many reps | Random data 1D | y=x | 5 0 | 3 -4 | -1 4 | 3 5 | ~50% | ~50% | ok |
| 4b | Test false positive rate on null data, High-D, many reps | Random data High-D | y=x | -3 0<br>2 -3<br>3 -2<br>⋮ ⋮ | -1 1<br>-4 0<br>0 2<br>⋮ ⋮ | 2 3<br>5 -2<br>4 5<br>⋮ ⋮ | -2 2<br>1 -3<br>0 -4<br>⋮ ⋮ | ~50% | ~50% | ok |

# 4 Test cases components

## 4.1 Intention and the initial expectation

In most conditions, the intention behind a test case is to test if the outcome of an analysis on a specific experimental design agrees to the expectation of the experimenter. While these expectation are often implicit, performing SAA often helps to make them explicit. Expectations might arise from intuitions about how a variable is supposed to influence the final result (e.g. an experimental manipulation should lead to a measurable effect when there is one), about the expected result of a given statistical measure (e.g. counterbalancing for removing confounding influences), an assumption on the generation of an experimental design (e.g. that every task condition occurs equally often), etc. The expectation can either be that the outcome has a *fixed* value (i.e. will always be the same), or that it comes from a distribution with defined properties (e.g. in the case of statistical control). Whenever possible, the expectation should include a clear description which result would confirm the expectation and which would violate it. In many cases violations of expectations can be obvious, as was the case in the initial example. In other cases it might be more difficult to detect a violation (see 4.7).

## 4.2 Diagnostic information

Diagnostic information is a variable that participates in generating pseudo-data. While in principle any information can be used as diagnostic information, in the following we will list typical examples. First of all, any design variable from the experimental design can be used as diagnostic information for SAA. This includes standard design variables such as the experimental condition, but also hidden design variables such as trial number or stimulus repetition. A second and related type of variable are covariates that are not part of the original experimental design, such as age, gender, IQ, or personality scores. Third, control data such as error rates, response times, pupil responses etc. can be interesting variables to test with SAA. Particularly noteworthy is to time-shift variables to detect temporal or sequence effects (e.g. always using the value of the previous trial). Finally, some diagnostic information can be used to perform sanity checks of the analysis pipeline. One such piece of diagnostic information is null data[1]. Another one is "**1**-data" (i.e. using the value 1 for each trial), which helps to detect basic internal problems of the pipeline such as biases towards one class in classification. The same diagnostic information can be used as input to multiple test cases (see main manuscript), e.g. to test the effect of different mapping functions, different experimental ranges, or different analysis methods, and one test case can have one or multiple sources of diagnostic information as input.

## 4.3 Analysis range: Selecting part of the experimental pipeline to check

Different SAA test cases can test the behaviour of different ranges (parts, units) of the experimental pipeline, from a single analysis step to multiple steps to the entire processing pipeline. Examples for analysis ranges in MVPA are cross-validated decoding, group-level analyses on decoding statistics (e.g. *t*-test on accuracy across subjects), or feature estimation (e.g. the generation of beta estimates from a BOLD time series).

## 4.4 Mapping functions: Generating pseudo-data from diagnostic information

For each variable that is tested with SAA, we need to generate data *y'* that only depends on the selected diagnostic information, but not on any other information, so that it is possible to uniquely ascribe the final result to the influence of the mapped diagnostic information. We do so by calculating

$$y'_i = f_i(x_i)$$

where $f_i$ is a suitable mapping from the selected diagnostic information $x_i$ to potential data. This transformation is a crucial step for SAA. The choices that primarily determine the mapping function are the analysis range, the analysis method, and assumptions about how the selected pieces of diagnostic information influence the data.

---

[1] To make simulations comparable to real analyses, the dimensionality of simulations should match the intrinsic dimensionality of the recorded data. The intrinsic dimension of the data is the "effective" number of variance components, and can e.g. be visually assessed by plotting the eigenvalues of the covariance matrix of the data – a scree plot – or defined numerically using properties of the eigenvalue spectrum (see e.g. Wackermann & Allefeld, 2009, p. 202f). The intrinsic dimensionality of data is typically not the dimension of the recording channels (e.g. fMRI voxels or EEG electrodes), but is often lower, but can also be higher (e.g. when multiple recording time points are combined, e.g. in time-frequency analyses).

Statistically speaking, a mapping functions implements a statistical model how the diagnostic information influence the measured data. The analysis range determines the format of the generated data. For example, testing the entire analysis pipeline of an fMRI experiment requires creating artificial BOLD time series, but for analyses that work on individual trial estimates (such as cross-validated decoding in the initial example) it can suffice to use the values of a design variable as data directly. Two simple but common mapping functions are:

1. *Identity f(x) = x:* A common assumption in neuroimaging is that a stimulus-responsive neuronal population will exhibit a linear relation between input x and measurements y (e.g. the more light the higher the neuronal response). The identity function y' = x can be used in such cases if the analysis method is not sensitive to scaling effects.
2. *Categorical f(x) = [$x_i$=1; else 0]$_n$:* Another widespread assumption is that different stimuli evoke different patterns of brain activity. These might be spatially distinct such as different regions encoding houses and faces (Kanwisher et al., 1997), or they might overlap, like for orientation tuning in V1 (Hubel & Wiesel, 1962, 1968). Diagnostic information that encodes categorical information (e.g. a design variable that codes the target stimulus category "house" as "1" and "face" as "2") can be recoded using dummy coding from standard statistics. For example[2], if a variable contains four categories encoded by the values 1 to 4, *x* = 2 will be mapped to *y'* = [0 1 0 0].

Other examples for mapping functions to generate data to test ranges that need temporally extended data (e.g. BOLD or EEG data) are the standard convolutions of onset and duration vectors with a canonical hemodynamic response function, which are routinely used as input to the general linear model in fMRI analyses. Spatial and/or temporal autocorrelations in the data can be added by employing additional spatial and/or temporal convolution (see e.g. Schreiber & Krekelberg, 2013). Such data can also be useful to create test cases for representational similarity analysis (RSA; Kriegeskorte et al., 2008).

### 4.5   Analysis
The analysis to use in a SAA test case is directly dictated by the SAA core, the "same analysis principle": The *same* analysis that should be used as in the main analysis. This includes using the same code, the same parameters, the same trials that go into the analysis, etc. (where this is not possible, the analysis should be kept as similar as possible). If different analyses should be run on the recorded data, each analysis needs to be tested with different test cases.

### 4.6   Outcome measures
There are numerous types of problems that can be detected using SAA, and these types depend on the chosen outcome measure that determine the result. But as for the analysis: There is not really anything to decide here, because the outcome measure is dictated by the intention, the analysis, and the selected range. For example, when testing cross-validated decoding, then primary outcome measures are classification accuracy or area under the receiver operating characteristic curve (AUC). When using SAA to test whether statistical outcomes are valid, then p-values, proportion of statistically significant results or range of confidence intervals could be used. When using SAA with artificially generated data, Monte-Carlo simulations can test whether the empirical alpha level matches the intended alpha level, e.g. whether indeed no more than 5% of all null simulations will be decided significant for α = 0.05. Finally, for testing general distributional assumptions, the mean or shape of the null could be tested. These are merely examples of typical outcome measures that can be used with SAA.

### 4.7   Comparing initial expectations and results
Once the result of a SAA test case is calculated, it is finally compared to the initial expectation of the experimenter (if such an expectation exists, which is indeed often the case). In general, we suggest using both (i) computational methods (that may or may not include statistical assessment) and (ii) visualization of the outcomes to test whether the outcome matches the expectation, because both complement each other[3]. One

---

[2] More generally, if *n* different values exist, each value *i* will be mapped to an *n*-dimensional vector containing a 1 at dimension *i* and 0 elsewhere.

[3] Humans can detect unexpected patterns in data because they rely on (often implicit) intuitions when thinking about data. Here, visualization helps to update intuitions and to detect deviations from intuitions. However, because humans are prone

example for computational methods that do not require statistical assessment would be decode the variable of interest from itself, i.e. using the same data as input and output, because should *always* yield an accuracy of 100% and should be significant. Another example would be performing a *t*-test on a perfectly counterbalanced experiment, because this should *always* yield the *t*-value $t = 0$ and should not be significant. In contrast, test cases that include stochastic influences at some point require statistical assessment of outcomes, e.g. those that use statistical control during the design or use null data as diagnostic information to test that not (significantly) more than $\alpha \times 100\%$ of the outcomes are reported significant. Note that statistical assessment can of course itself give a false positive, so not every significant outcome is necessarily alarming (indeed, some significant deviations are expected if multiple SAA tests are performed which requires adjustment for multiple testing, see main paper).

---

to over-interpreting patterns in data and might miss differences that look small on the visualisation, employing statistical methods to test clearly defined assumptions is important as well.

# 5 Additional fMRI results for empirical example

Figures D.1-D.8 show the decoding analyses results for the variance confound example that include all pairwise combinations of 1st level regressors created with 3/6/12 left button presses (rows) and 3/6/12 right button presses (columns) per run, respectively, for all time bins between 0-16s (FIR bins 1-8). The diagonal in each plot (always 2x2 brain renderings belong together) shows the three confound-free analyses in which equally many button presses were available for the left and the right condition. Note again that the number of button presses only differed to create first level regressor images, but that the decoding analyses were all run on equally many examples for left and right (always 6 images per class, one per experimental run).

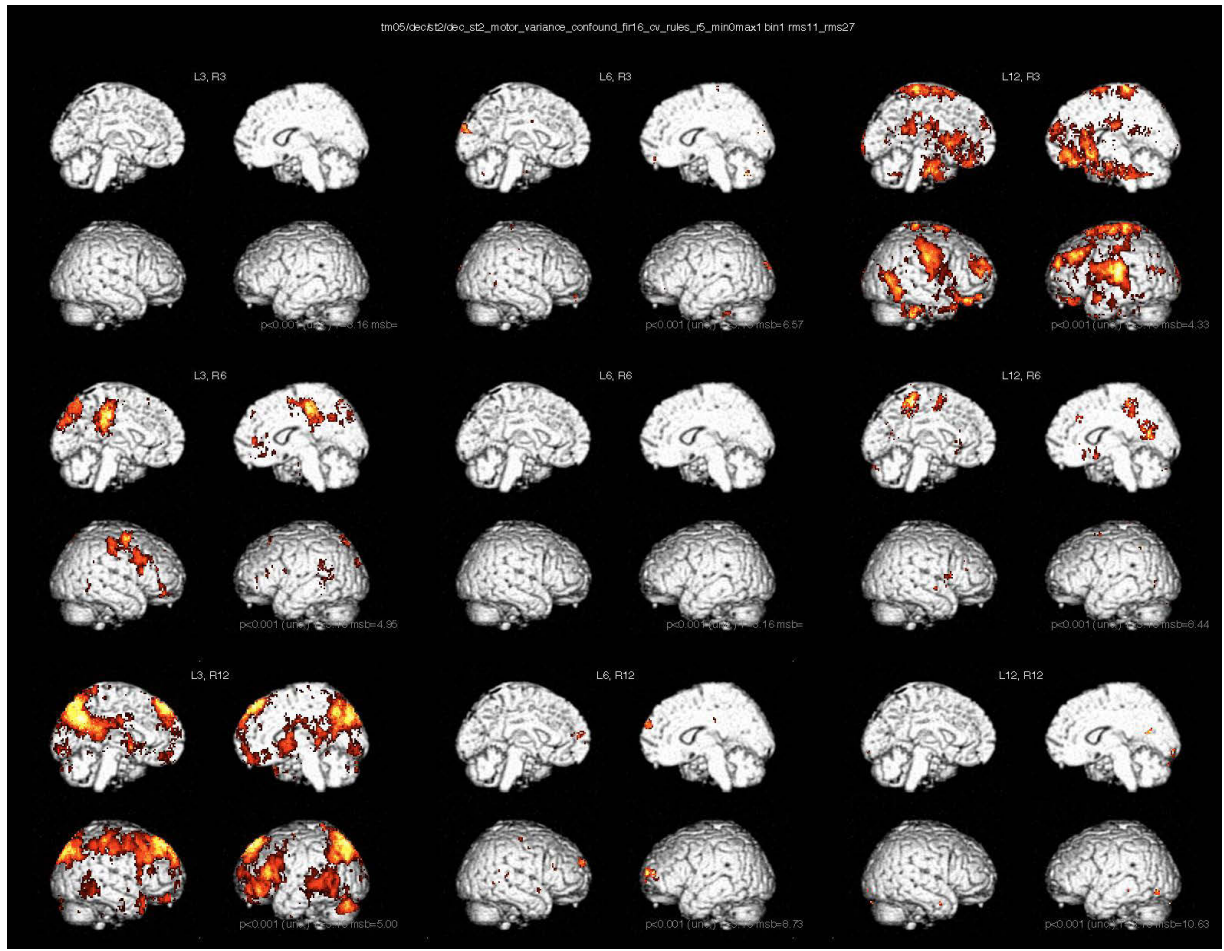

**Figure SI D.1. Bin 1 (0-2s).** Decoding analyses for the variance confound example that include all pairwise combinations of 1st level regressors created with 3/6/12 left button presses (rows) and 3/6/12 right button presses (columns) per run

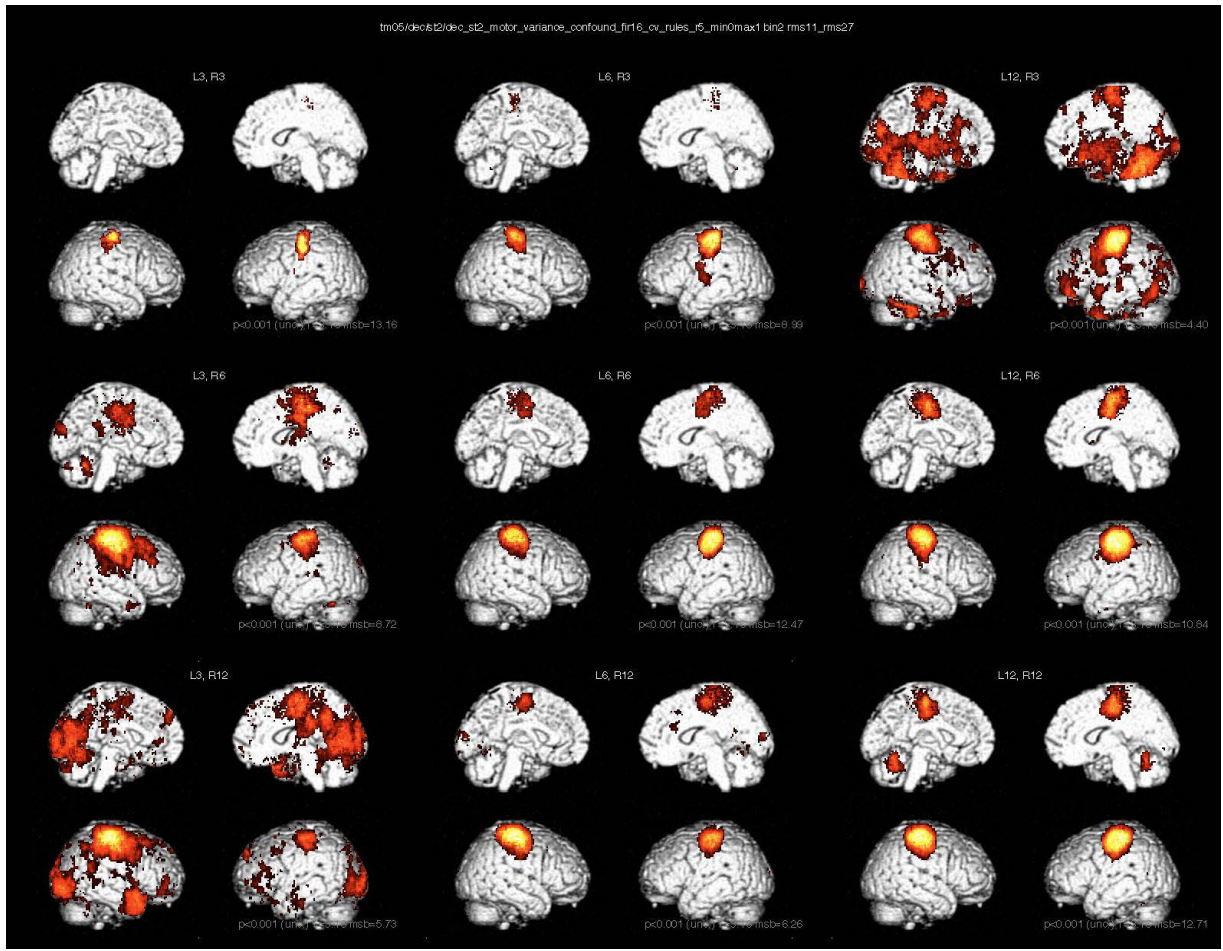

**Figure SI D.2. Bin 2 (2-4s).** Decoding analyses for the variance confound example that include all pairwise combinations of 1st level regressors created with 3/6/12 left button presses (rows) and 3/6/12 right button presses (columns) per run

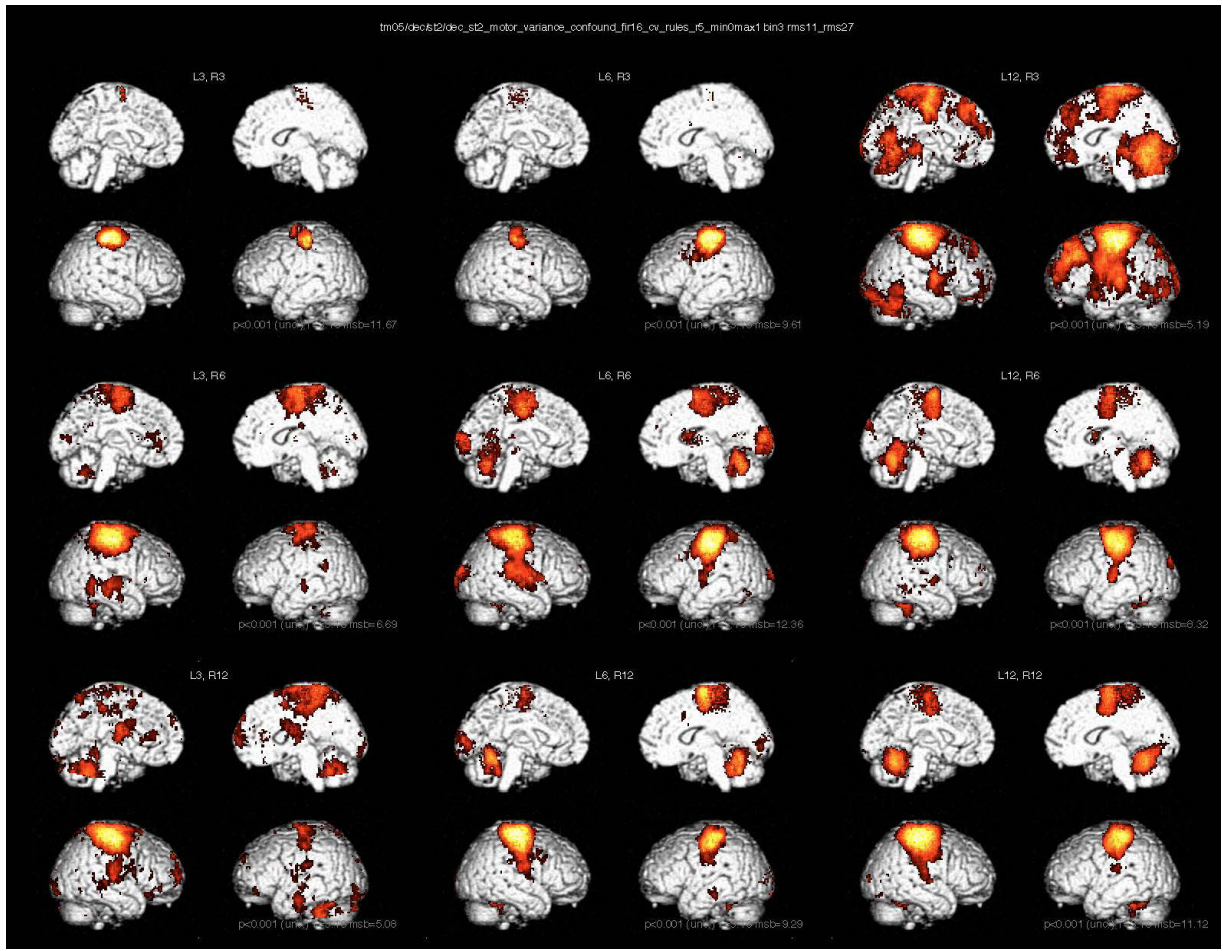

**Figure SI D.3. Bin 3 (4-6s).** Decoding analyses for the variance confound example that include all pairwise combinations of 1st level regressors created with 3/6/12 left button presses (rows) and 3/6/12 right button presses (columns) per run

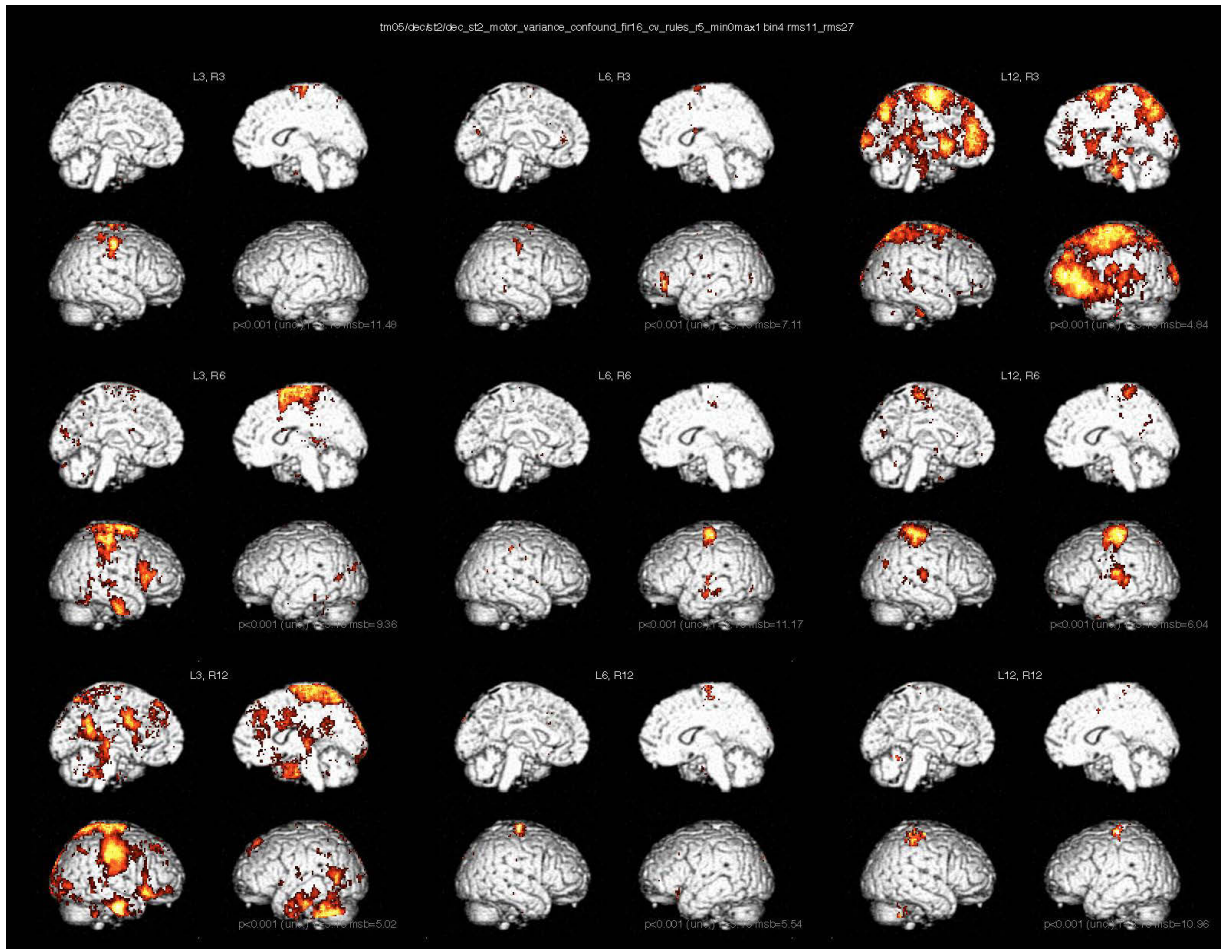

**Figure SI D.4. Bin 4 (6-8s).** Decoding analyses for the variance confound example that include all pairwise combinations of 1st level regressors created with 3/6/12 left button presses (rows) and 3/6/12 right button presses (columns) per run

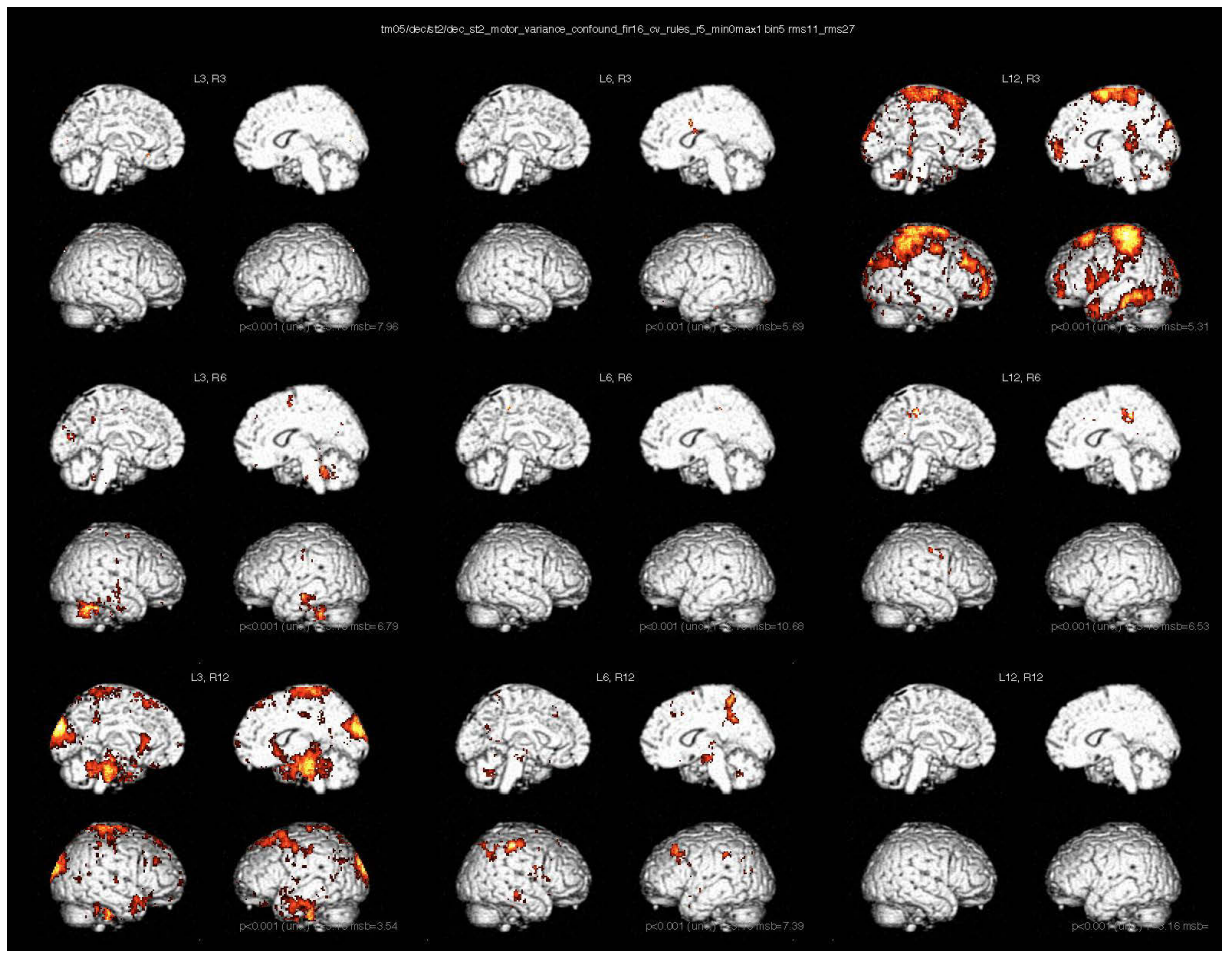

**Figure SI D.5. Bin 5 (8-10s)** Decoding analyses for the variance confound example that include all pairwise combinations of 1st level regressors created with 3/6/12 left button presses (rows) and 3/6/12 right button presses (columns) per run

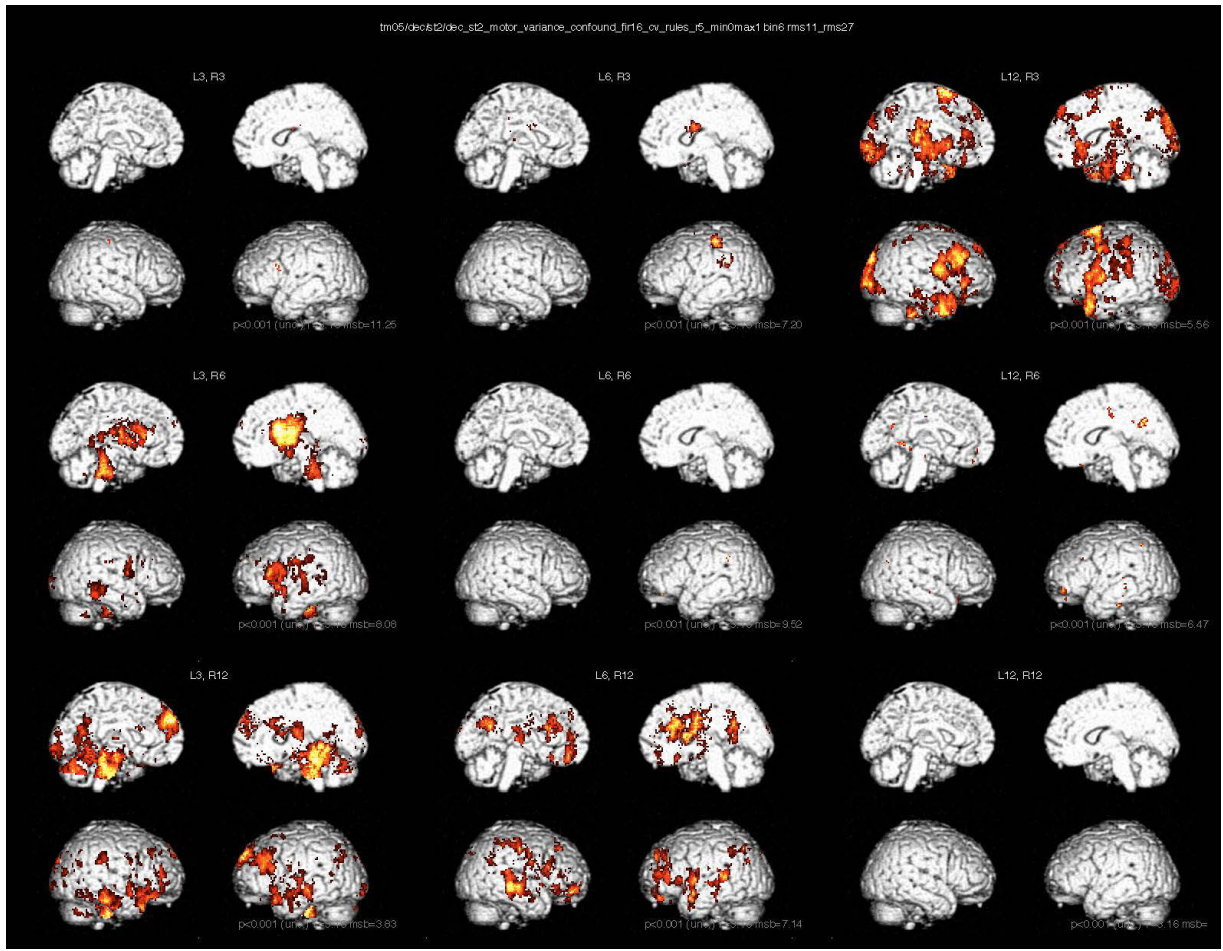

**Figure SI D.6. Bin 6 (10-12s).** Decoding analyses for the variance confound example that include all pairwise combinations of 1st level regressors created with 3/6/12 left button presses (rows) and 3/6/12 right button presses (columns) per run

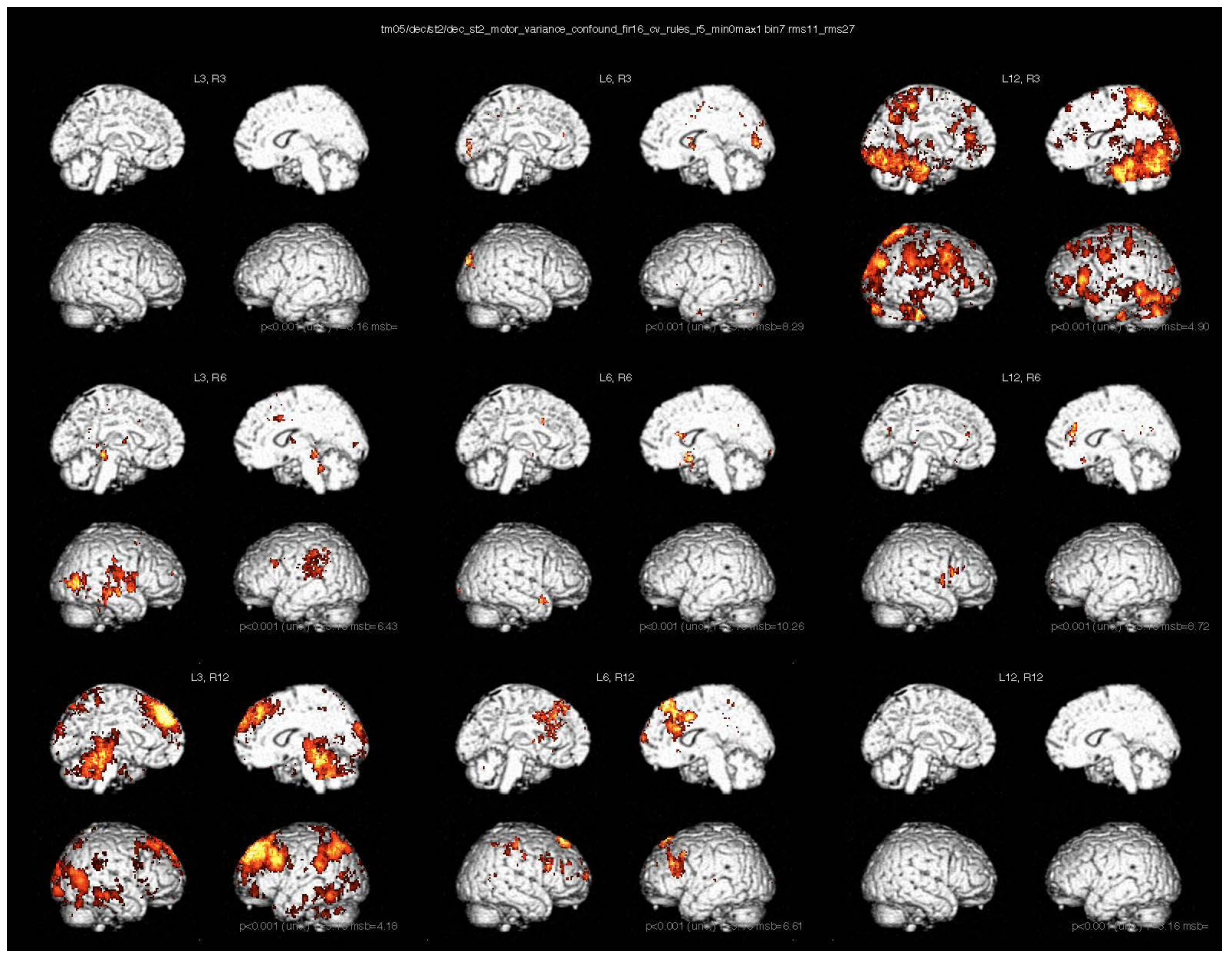

**Figure SI D.7. Bin 7 (12-14s).** Decoding analyses for the variance confound example that include all pairwise combinations of 1st level regressors created with 3/6/12 left button presses (rows) and 3/6/12 right button presses (columns) per run

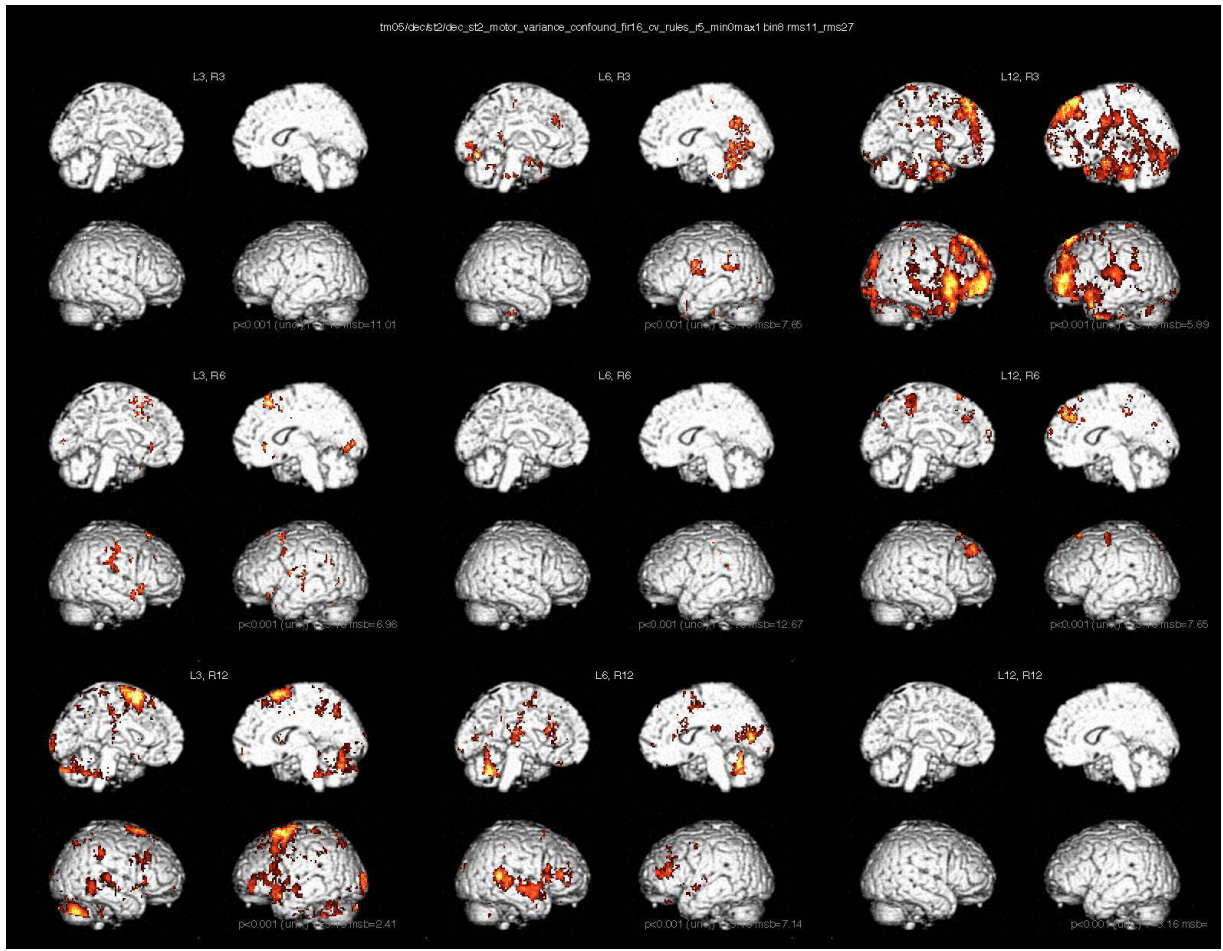

**Figure SI D.8. Bin 8 (14-16s).** Decoding analyses for the variance confound example that include all pairwise combinations of 1st level regressors created with 3/6/12 left button presses (rows) and 3/6/12 right button presses (columns) per run

# 6 SAA setup

Here we explain how the SAA analysis of the main paper has been conducted in more detail. This might also help researchers that what to implement SAA for their study.

## 6.1 Data extraction & dummy coding

The first step is to extract the data for each trial. In the main neural analysis, we need to create a list with onset and length of each event for each condition that we are interested in. For example, we might have Left and Right button presses in one run in the trials trials(L): {3, 4, 7, 12} and trials(R): {1, 2, 6, 13}, and these might occur at onsets(L): {17, …} and onset(R): {24, …} seconds after trial onset.

In parallel, for the SAA analysis we create lists for each condition that contain information about each variable that we want to analyse. For example, the "task condition" variable lists for this run might look like: TaskCondition(L): {A, D, C, B}, TaskCondition(R): {C, D, A, B}. See Table SI 2 below for a full coding example.

Because the "task condition" variable is *categorical*, we need to perform dummy-coding as an additional step, because there is no intrinsic ordering of the conditions A-D. We do so by creating a 4-dimensional binary representation [a b c d], where always the current task condition is 1 and all the others are 0 (e.g. a trial of condition "A" would be represented as [1 0 0 0], "B" as [0 1 0 0], etc.).

**Table SI 2 – SAA coding for a single example run.**

| Predictors | Left Button Press Trials, Run 1 | | | | Right Button Press Trials, Run 1 | | | |
|---|---|---|---|---|---|---|---|---|
| Trialnr | 3 | 4 | 7 | 12 | 1 | 2 | 6 | 13 |
| Decoding Class | L | L | L | L | R | R | R | R |
| Onset | 17 | | … | | 24 | | … | |
| TC (Task Condition) | A | D | C | B | C | D | A | B |
| … | | | | | | | | |
| Constant | 1 | 1 | 1 | 1 | 1 | 1 | 1 | 1 |
| Randn 1 | 0.7 | | | | | | | |
| Randn 2 | -0.2 | | | | random numbers from N(0, 1) | | | |
| … | 1.2 | | | | | | | |
| Randn n | -0.8 | | | | | | | |
| Dummy coding | | | | | | | | |
| TC.A | 1 | 0 | 0 | 0 | 0 | 0 | 1 | 0 |
| TC.B | 0 | 0 | 0 | 1 | 0 | 0 | 0 | 1 |
| TC.C | 0 | 0 | 1 | 0 | 1 | 0 | 0 | 0 |
| TC.D | 0 | 1 | 0 | 0 | 0 | 1 | 0 | 0 |
| Trial summaries | | | | | | | | |
| Mean(trialnr) | | 6.5 | | | | 5.5 | | |
| Mean(TC=A) | | 0.25 | | | | 0.25 | | |
| Mean(TC=B) | | … | | | | | | |
| … | | | | | | | | |

## 6.2 Run-wise summaries

In this example, decodings are performed on run-wise beta estimates rather than on individual trials. We thus need to parallel this step for our SAA tests. This step can be omitted in studies that decode on single trial data directly (another popular approach in neuroimaging). Different motivations underlie calculating 1$^{st}$ level regression estimates ("betas"): The main ones are de-mixing of the voxel time series caused by overlapping effects of different factors, de-convolution of the HRF (often called "temporal compression"), and linear averaging for noise-reduction. Because in SAA we want to test whether variables that are *independent* covariates in the main analysis could cause effects in the *dependent* measured variable of the main analysis, the only motivation that we need to parallel is linear averaging. Thus, we can parallel the beta regression estimation by calculating run-wise means for each signal. (If we had calculated other measures than linear means, or if we want to test the hypothesis that other than linear influences of the independent variables cause differences in brain activity (e.g. deviation from the mean), we might want to calculate also other summary measures such as the variance or the like here.)

## 6.3 SAA: Multiple comparison correction in result displays

As visual aid, we show second level outcomes in a graded scheme to not be confused too fast. Variables can either be Bonferroni significant, significant, or show a trend. To also detect potential malicious below chance influences we also display these, but not colour-coded like the variables that could cause false-positive results. All variables that are not random (class 1 side, ntrials, and const, class 2) are not included to calculate the correction (because they are not random processes), and the correction is only calculated for the variables that are indeed displayed. The whole purpose of this procedure is to easier detect potentially harmful confounds, because of course a variable might still be confounding even if it is not detected here.

## 6.4 Suggested classes for SAA tests

To ease interpretation, we suggest to sort SAA tests into four classes that seem typical for decoding studies: "Sanity checks" (Class 1), "guaranteed outcome" (Class 2), "average outcome" (Class 3), and "control data" (Class 4). A fifth class that is mentioned below which we haven't included in the analysis here are variables that can be included for unit testing. The classes are ordered by consecutive influences: Confounds of variables with smaller class numbers will probably also influence variables with larger class numbers, but not vice versa. So problems in classes with smaller numbers should be solved first.

*Class 1:* The first class of variables contains sanity checks. If these already fail, everything else will probably fail, too. In the example, we have added three such checks: First, if we can decode button press side if our data is button press side. This should yield a decoding accuracy of 100% in each and every single case, and if it doesn't, there seems to be a severe problem in the code. Second, we included a constant regressor that is always "1" for each trial. Here, decoding on the mean (="1") should result in 50% accuracy in our case, because we used leave-one-pair-out cross-validation. Decoding on the sum might be ok if everything else seems fine, because we had an unequal number of trials but we averaged them, so a significant result here should not bother us too much. Finally, this class also contains the 10,000 'randn' variables that contained random data for each trial. Because we used a significance level of α=5%, single significant outcomes are not interesting for us, but only if significantly more than 5% of the 10,000 tests return significant results.

*Class 2:* The next class contains design variables that should *always* yield a certain outcome, in each and every single experiment. An example would be a variable that we had balanced against the button press variables, e.g. were we had equally many left and right button presses for level of that variable. Here, decoding should **never** be possible using left and right button presses. Thus, an alarming outcome for such a variable would be any deviation from chance level, not only if the 2$^{nd}$ level t-test result is significant. No such variable is included in the current example, but it typically is in many experiments.

*Class 3:* The outcome of other design principles might not be as strict, but should guarantee that **on average** the design variables are unbiased. All design principles that we used in our example fall into this category, because they all have been (pseudo-)randomized. The interpretation of the results of variables in this class differs between design creation and confound detection during data analysis. During the design creation stage, many designs should be created to ensure that the creation process is valid *on average* (see 'randn' analysis above). During the data analysis stage, we want to check whether the *concrete* design that was randomly drawn for actual data collection is ok.

An idea that easily comes into mind here is *design selection*, i.e. not to take any random design but to continue creating designs until one is found that has no confounds at all *before* collecting data. However, a couple of critical issues need to be considered here: First, it is often impossible to generate designs so that every variable is completely confound-free. Second, selecting designs based on certain criteria reduces (often drastically) the space from which designs are drawn, and thus might introduce new confounds in variables that have not been used for selection. Thus, one might not want to generate the best possible design but just to throw out designs that are really bad. Finally, when using a design selection procedure, it is important to make sure that the designs that are selected by this procedure are themselves unbiased on average, and then to only use a *random* design from this pool of allowed designs; otherwise one runs into the problem of "double dipping" (Kriegeskorte et al., 2009).

***Class 4:*** The final class in the example SAA analysis are the ***control variables*** 'reaction time', 'time button of button press', and 'answer correct or wrong'. Control data is routinely analysed in MVPA studies, but is typically analysed with different methods than the main data of interest, potentially causing to both overlook potential confounds because the other method is not sensitive to it (or vice versa).

***Class 5:*** A final class that we did not include in this example analysis but that is probably helpful to detect confounds (or bugs) in a design is to include variables for ***unit testing***. In general, one can add any variable in this class. The only difference to the previous classes is that one does not really expect confounds in these variables, and thus the false positive control should be much more liberal. Specifically, we suggest to calculate separate multiple comparison corrections for this class and all other classes, so that the results of the analyses of other classes stay the same independent of how many unit testing analyses one conducts. In general, multiple comparison correction in SAA results displays only works as a visual aid, not as statistical tests.

# 7 Variance confound decoding in simulations (Nearest Centroid and linear SVM)

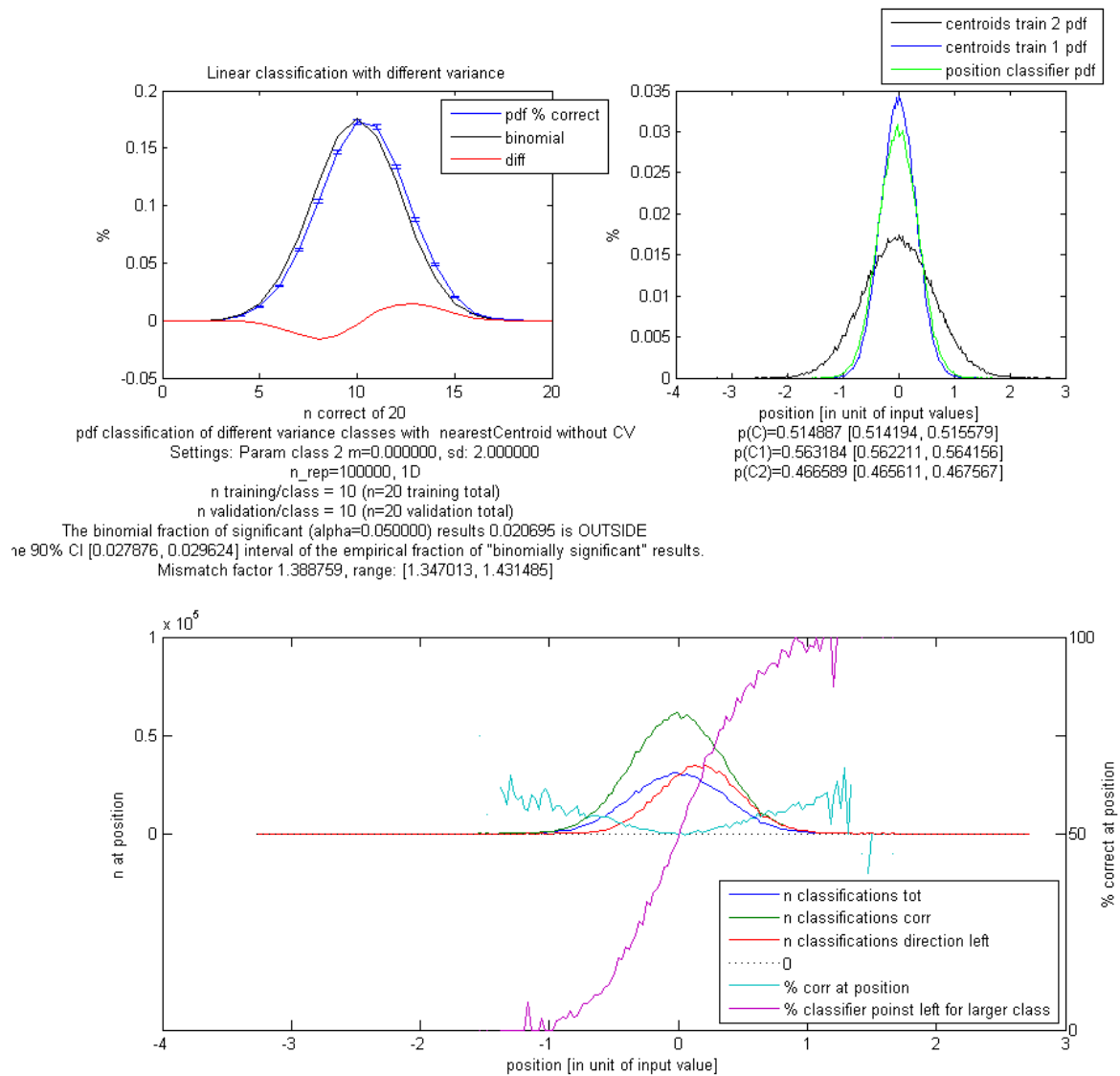

**Figure SI E.1. Variance decoding with a nearest-centroid classifier.** Scenario: 10-fold Leave-one-pair-out cross-validation with 20 random 1d numbers per dataset, 100,000 simulations. The data of both conditions is drawn from two normal distributions, that have the *same mean* and but differ in their variance (class1: *N*(m=0, sd=1); class 2: *N*(m=0, sd=2)). *Top left:* Occurrence probability of all possible n correct (from 0/20 to 20/20), results in simulation (blue) deviate systematically (red) from the binomial distribution (black), including a shift of the mean (p(C)=0.51 instead of .50; shown below right plot). As predicted, this is driven by predicting Class 1 (with smaller variance) higher (p(C1)=0.56, +6% from chance), the misclassifying class 2 (p(C2)=0.47, -3% from chance). *Top right*: Distribution of data centroids for each class (blue, black; centroids are the means of the training data) and the positions of the NN-classifiers (green). Although the classifier peaks at 0, it also occurs at both sides. *Bottom:* Additional insights in the classifier behaviour at different positions. Most interesting should be % correct at different positions (cyan), that is on average 50% for classifiers at position 0 but increases (as predicted) if the classifier is positioned left or right. Integrating over all positions thus results in the observed above chance decoding accuracy. The purple line shows the average direction of the classifier at each position (fraction of pointing to the left).

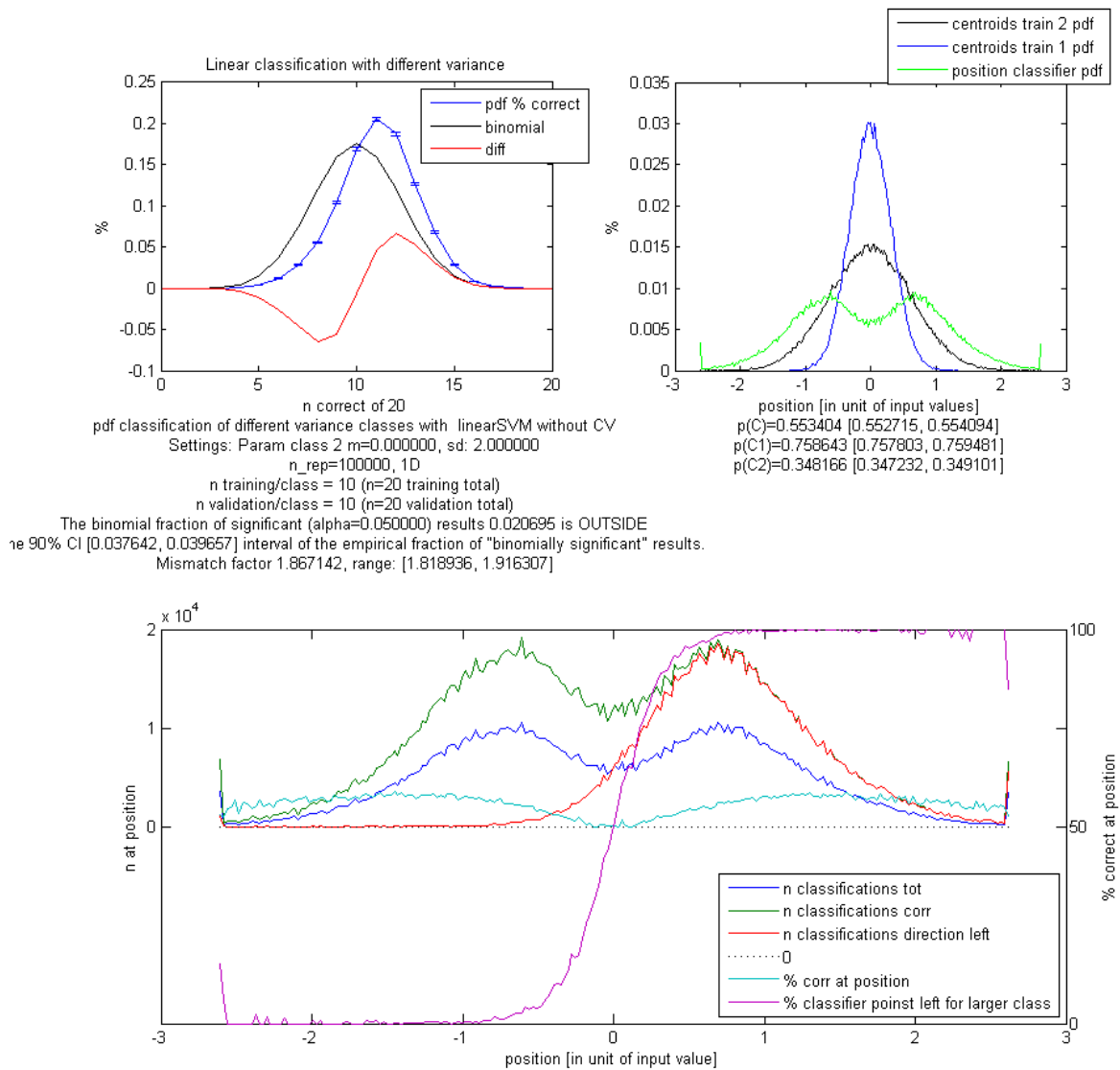

**Figure SI E.2. Above chance decoding with a linear SVM classifier.** Setting like in figure E.1 above. The SVM can distinguish the classes even better (p(C)=0.55 correct) because the classifier has a wider spread and more frequently occurs outside the middle (green curve in upper right and lower plot).

# 8 Supplemental References